\documentclass[journal=jacsat,manuscript=article]{achemso}
\usepackage[version=4]{mhchem} 
\usepackage[T1]{fontenc}       
\setkeys{acs}{articletitle = true}
\usepackage[labelfont=bf,font=small,labelsep=period,justification=raggedright]{caption}

\usepackage{changepage}

\usepackage{amsmath,amssymb}

\usepackage{mathtools}

\usepackage[utf8]{inputenc}

\usepackage{units, xfrac}

\allowdisplaybreaks

\SectionNumbersOn 

\usepackage{nameref}
\usepackage{hyperref}

\usepackage{enumitem}
\usepackage{graphicx}
\usepackage{amssymb}

\usepackage{mhchem}

\usepackage[euler]{textgreek}

\usepackage{color}
\usepackage[table,xcdraw,dvipsnames]{xcolor}

\def\be{\begin{equation}}
\def\ee{\end{equation}}

\newcommand{\koff}{k_\mathrm{off}}

\newcommand{\vR}{v_\mathrm{R}}
\newcommand{\vW}{v_\mathrm{W}}

\newcommand{\kon}{k_\mathrm{on}}
\newcommand{\konW}{k_\mathrm{on}^\text{W}}
\newcommand{\konR}{k_\mathrm{on}^\text{R}}
\newcommand{\konA}{k^{\mathrm{A}}_\mathrm{on}}
\newcommand{\konI}{k^{\mathrm{I}}_\mathrm{on}}

\newcommand{\kf}{k_\mathrm{f}}
\newcommand{\kb}{k_\mathrm{b}}
\newcommand{\knet}{k_\mathrm{net}}

\newcommand{\kfud}{k^{\mathrm{u} \rightarrow \mathrm{d}}_\mathrm{f}}
\newcommand{\kfdu}{k^{\mathrm{d} \rightarrow \mathrm{u}}_\mathrm{f}}

\newcommand{\kfudL}{k^{\mathrm{u} \rightarrow \mathrm{d, L}}_\mathrm{f}}
\newcommand{\kfduL}{k^{\mathrm{d} \rightarrow \mathrm{u, L}}_\mathrm{f}}

\newcommand{\dFud}{\Delta F_{\mathrm{u} \rightarrow \mathrm{d}}}

\newcommand{\dFdu}{\Delta F_{\mathrm{d} \rightarrow \mathrm{u}}}

\newcommand{\lonA}{\ell^{\mathrm{A}}_\mathrm{on}}

\newcommand{\loffA}{\ell^{\mathrm{A}}_\mathrm{off}}

\newcommand{\Vup}{V_{\mathrm{u}}}
\newcommand{\Vdown}{V_{\mathrm{d}}}

\newcommand{\kB}{k_\mathrm{B}}

\let\mpar=\marginpar
\renewcommand\marginpar[1]{\mpar{\raggedright \scriptsize #1}}

\usepackage{setspace}

\newcommand{\koffW}{k_\mathrm{off}^{\mathrm{W}}}
\newcommand{\koffR}{k_\mathrm{off}^{\mathrm{R}}}

\author{Vahe Galstyan}
\affiliation{Biochemistry and Molecular Biophysics Option, California Institute of Technology, Pasadena, California 91125, United States}
\author{Rob Phillips}
\affiliation{Department of Physics, California Institute of Technology, Pasadena, California 91125, United States}
\alsoaffiliation{Department of Applied Physics, California Institute of Technology, Pasadena, California 91125, United States}
\alsoaffiliation{Division of Biology and Biological Engineering, California Institute of Technology, Pasadena, California 91125, United States}
\email{phillips@pboc.caltech.edu}
\phone{(626) 395-3374}

\title{Allostery and Kinetic Proofreading}

\begin{document}

\maketitle
\singlespacing

\newpage
\begin{abstract}
Kinetic proofreading is an error correction mechanism present in the processes of the central dogma and beyond,
and typically requires 
the free energy of
nucleotide hydrolysis for its operation.
Though the molecular players of many biological proofreading schemes are known, 
our understanding of how
energy consumption is managed to promote fidelity remains incomplete.
In our work, we introduce an alternative conceptual scheme called ``the piston model of proofreading'' where enzyme activation 
through hydrolysis is replaced with allosteric activation achieved through 
mechanical work performed by a piston on regulatory ligands.
Inspired by Feynman's ratchet and pawl mechanism, we consider a mechanical engine designed 
to drive the piston actions powered by a lowering weight,
whose function is analogous to that of ATP synthase in cells.
Thanks to its mechanical design, the piston model allows us to tune the ``knobs'' of the driving engine and
probe the graded changes and trade-offs between speed, fidelity and energy dissipation.
It provides an intuitive explanation of the conditions necessary for optimal proofreading 
{\color{black}and
reveals the unexpected capability of allosteric molecules to beat the Hopfield limit of fidelity 
by leveraging the diversity of states available to them.
The framework that we built for the piston model can also serve as a basis for additional studies
of driven biochemical systems.}
\end{abstract}

\section{Introduction}

Many enzymatic processes in biology need to operate with very high fidelities in order to ensure the physiological well-being of the cell.
Examples include the synthesis of molecules making up Crick's so-called ``two great polymer languages'' 
(i.e. replication, \cite{Kunkel2004} transcription, \cite{SydowCramer2009} and translation \cite{Rodnina2001}), as well as processes that go beyond those of the central dogma, such as
protein ubiquitylation mediated by the anaphase-promoting complex, \cite{LuKirschnerScience2015} 
signal transduction through MAP kinases, \cite{SwainBiophysJ2002}
pathogen recognition by T-cells,\cite{Mckeithan1995, Goldstein2004} or protein degradation by the 26S proteasome. \cite{Bard2019}
In all of these cases, the designated enzyme needs to accurately select its correct substrate from a pool of incorrect substrates.
Importantly, the fidelity of these processes that one would predict solely based on the free energy difference between correct
and incorrect substrate binding is far lower than what is experimentally measured, 
raising a challenge of explaining the high fidelities that this naive equilibrium thermodynamic thinking fails to account for.

The conceptual answer to this challenge was provided more than 40 years ago in the work of John Hopfield \cite{Hopfield1974}
and Jacques Ninio \cite{NinioBiochimie1975} and was coined ``kinetic proofreading'' in Hopfield's elegant paper entitled 
``Kinetic Proofreading: A New Mechanism for Reducing Errors in Biosynthetic Processes Requiring High Specificity.'' \cite{Hopfield1974}
The key idea behind kinetic proofreading is to introduce a delay between substrate binding and turnover steps,
effectively giving the enzyme more than one chance to release the incorrect substrate (hence, the term ``proofreading'').
The sequential application of substrate filters on the way to product formation gives directionality to the flow of time and
is necessarily accompanied by the expenditure of free energy, making kinetic proofreading an intrinsically nonequilibrium
phenomenon.
In a cell, this free energy is typically supplied to proofreading pathways through the hydrolysis of energy-rich nucleotides,
whose chemical potential is maintained at large out-of-equilibrium values 
through the constant operation of the cell's metabolic machinery
(e.g. the ATP synthase).

Since its original formulation by Hopfield and Ninio, the concept of kinetic proofreading has been generalized and
employed in explaining many of the high fidelity processes in the cell.
\cite{MuruganPNAS2012, WongPRE2018, SartoriPRL2013, Bard2019, DepkenCellRep2013, QianAnnuRev2007, SemlowTrends2012, BanerjeePNAS2017}
However, despite the fact that the molecular players and mechanisms of these processes have been largely identified,
we find that 
an intuitive picture of how energy transduction promotes biological fidelity is still incomplete.
To complement our understanding of how energy is managed to beat the equilibrium limit of fidelity,
we propose a conceptual model called ``the piston model of kinetic proofreading'' where 
chemical hydrolysis is replaced with mechanical work performed by a piston on an allosteric enzyme.
Our choice of allostery is motivated by the fact that in proofreading schemes hydrolysis typically triggers 
a conformational change in the enzyme-substrate complex and activates it for product formation\cite{Blanchard2004, YanMarko2001,
Semlow2012} - 
an effect that our model achieves through the binding of a regulatory ligand to the enzyme.
By temporally controlling the concentration of regulatory ligands which determine the catalytic state of the enzyme,
the piston sequentially changes the enzyme's state from inactive to active, creating a delay in product
formation necessary for increasing the fidelity of substrate discrimination.
The piston actions are, in turn, driven by a Brownian ratchet and pawl engine powered by a lowering weight,
whose function is akin to that of ATP synthase.
The mechanical design of the piston model allows us to transparently control the energy input into the system by tuning
the ``knobs'' of the engine and examine the graded changes in the model's performance metrics, 
intuitively demonstrating the driving conditions required for optimal proofreading.

We begin the presentation of our results by first introducing in section~\ref{section:model} the high-level concept behind the piston 
model of proofreading, 
while at the same time drawing parallels between its features and those of Hopfield's original scheme.
Then, in sections~\ref{section:engine} and \ref{section:enzyme} we provide a comprehensive description of the two key constituents of the piston model, namely,
the Brownian ratchet and pawl engine that drives the piston actions, and the allosteric enzyme whose catalytic state
is regulated by an activator ligand.
This is following by building the full thermodynamically consistent framework of the piston model in section~\ref{section:coupling} where we
couple the external driving mechanism to the enzyme and introduce the expressions for key performance metrics of the model.
In the remaining sections~\ref{section:tradeoffs} and \ref{section:enzyme_tuning}, 
we explore how tuning the ``knobs'' of the engine leads to graded changes and
trade-offs between speed, fidelity and energy dissipation,
and probe the performance limits of the piston model as a function of a select set of key enzyme parameters.

\newpage
\section{MODEL}
\label{section:model}

{\color{black}
The piston model of kinetic proofreading is designed in analogy with Hopfield's scheme.
The main idea there was to give the enzyme a second chance to discard the wrong substrate by
introducing an additional kinetic intermediate for the enzyme-substrate complex (Figure 1A).
The difference between substrate binding energies in Hopfield's original formulation was based solely
on their unbinding rates (i.e. $\koffW > \koffR$ and $\konW=\konR = \kon$) -- a convention we adopt throughout our analysis.
The first layer of substrate discrimination in Hopfield's scheme is achieved during the initial binding event
where the ratio of right and wrong substrate-bound enzymes approaches $\koffW / \koffR$.
The complex then moves into its catalytically active high energy state accompanied
by the hydrolysis of an NTP
molecule, after which the second discrimination layer is realized.
Specifically, right and wrong substrates are turned into products with an additional bias given by
the ratio of their Michaelis constants, namely, $(\koffW + r)/(\koffR+r)$. 
Importantly, for this second layer to be efficiently realized, the rates of binding directly to the second
kinetic intermediate need to be vanishingly small in order to prevent the incorporation of 
unfiltered substrates.\cite{Hopfield1974}}

\begin{figure}[!ht]
	\centerline{
		\includegraphics[scale=1.00]{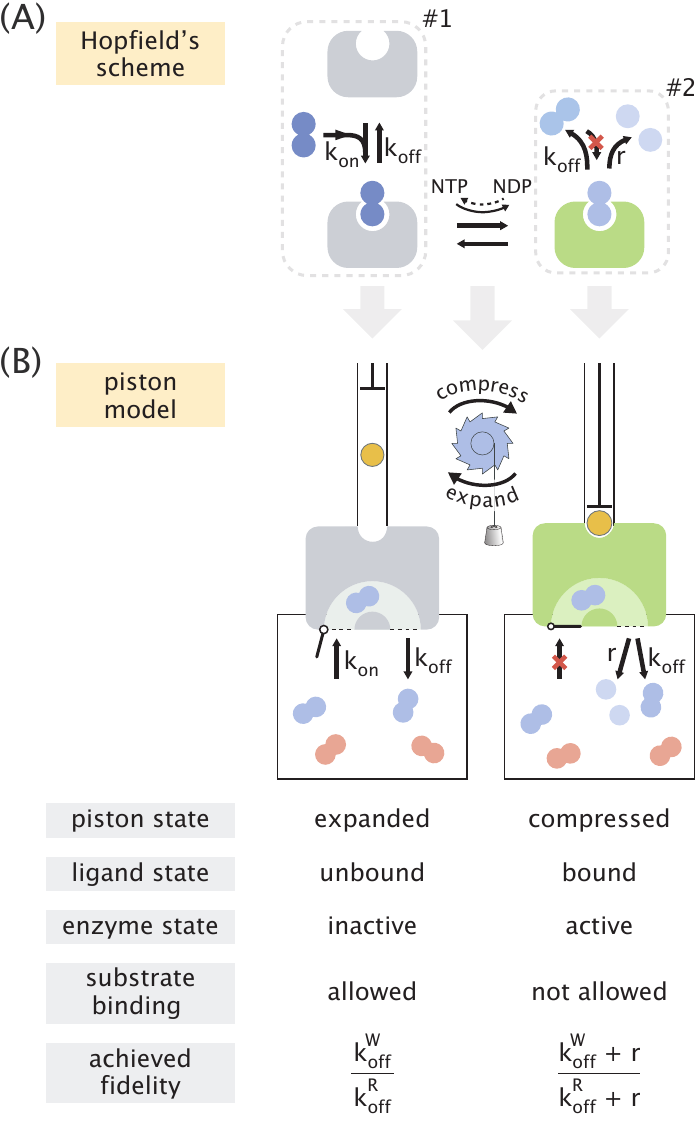}}
	\caption{Conceptual introduction to the piston model.	
	(A) Hopfield's scheme of kinetic proofreading where two layers of substrate discrimination take place on the driven pathway -- the first one during the initial binding of energy-rich substrates (\#1 in the diagram) and the second one upon the release of the energy-depleted substrates (\#2 in the diagram). Energy consumption takes place during the hydrolysis reaction $\text{NTP} \rightleftharpoons \text{NDP}$ accompanying the transition between the two intermediates.
	(B) Pedagogically simplified conceptual scheme of the piston model.
	The orange circle represents the activator ligand. Blue and red colors stand for the right and wrong substrates, respectively.
	The closed ``entrance door'' along with
	the red cross on the binding arrow in the active state of the enzyme
	suggests the vanishingly small rate of substrate binding
	when in this state.
	{\color{black}The ratchet with a hanging weight stands for the mechanical engine that drives the piston actions.}
	Various features of the system in the two piston states, 
	along with the expressions for achieved fidelities are listed below the diagram.
	Transparent arrows between panels A and B indicate the analogous parts 
	in Hopfield's scheme and the piston model.}
	\label{fig:piston_model_concept}
\end{figure}

With this information in mind, consider now the conceptual illustration of the piston model shown in Figure~\ref{fig:piston_model_concept}B,
{\color{black}
where we have made several pedagogical simplifications to help verbally convey the model's intuition, 
reserving the full thermodynamically consistent treatment to the following sections.}
The central constituent of the model is an allosteric enzyme, 
the catalytic activity of which is regulated by activator ligands (the orange circle).
{\color{black}
The enzyme is inactive 
when it is not bound to a ligand, and, conversely, it is active
when bound to a ligand.}
The volume occupied by ligands and hence, their concentration is, in turn, controlled by a piston.
The ligand concentration is 
very low when the piston is expanded, and very high when the piston is compressed
{\color{black}
in order to guarantee that 
in those piston states the ligand is free and bound to the enzyme, respectively.}

The active site of the enzyme is exposed to a container 
filled with right and wrong substrates of concentrations $[\text{R}]$ and $[\text{W}]$, respectively,
which we take to be equal for the rest of our analysis ($[\text{R}] = [\text{W}]$).
And unlike in Hopfield's scheme where the substrates exist in energy-rich and energy-depleted states
(e.g. tRNAs first arrive in the {EF-Tu$\cdot$GTP$\cdot$tRNA} ternary complex and then
release EF-Tu and GDP after hydrolysis), in the piston model substrates exist in a single state
and do not carry an energy source.
In the expanded piston state (Figure~\ref{fig:piston_model_concept}B, left), 
substrates can bind and unbind to the inactive enzyme, 
but do not get turned into products.
The highest level of discrimination achievable in  this state therefore becomes
\begin{align}
\eta_1 = \frac{\koffW}{\koffR},
\end{align}
in analogy to that achieved during the initial binding step of 
Hopfield's scheme.

After the first layer of substrate discrimination is established in the expanded state of the piston,
mechanical work is performed to compress it. This increases the ligand concentration, which,
in turn, leads to the activation of the enzyme where catalytic action is now possible.
To prevent the incorporation of unfiltered substrates, we assume that in the active enzyme state the rate of substrate
binding is vanishingly small, similar to Hopfield's treatment
(Figure~\ref{fig:piston_model_concept}B, right).
If the piston is kept compressed long enough, 
a filtered substrate that got bound earlier when the piston was expanded will experience one of these two outcomes: 
it will either turn into a product with a rate $r$ (which is taken to be the same for the two kinds of substrates) 
or it will fall off with a rate $k_\text{off}$.
The product formation reaction will take place with probability $r/(\koff + r)$. 
Thus,
due to the difference in the falloff rate constants between the right and the wrong substrates,
the extra fidelity achieved after piston compression equals 
\begin{align}
\eta_2 = \frac{\koffW + r}{\koffR+r}.
\end{align}
Once this extra fidelity is established, the piston is expanded back, repeating the cycle
(the detailed derivation of the results for $\eta_1$ and $\eta_2$ is provided in Supporting Information section A).
Notably, the total fidelity achieved during the piston expansion and compression cycle, namely,
\begin{align}
\eta = \eta_1 \eta_2 = \left( \frac{\koffW}{\koffR} \right) \left( \frac{\koffW + r}{\koffR+r} \right),
\end{align}
exceeds the Michaelis-Menten fidelity ($\eta_2$) by a factor of $\eta_1 = \koffW/\koffR$,
demonstrating the attainment of efficient proofreading.

The cyclic compressions and expansions of the piston in our model also stand in direct analogy to the
hydrolysis-involving transitions between the two enzyme-substrate intermediates in Hopfield's scheme.
{\color{black}In particular, they need to be externally driven for the mechanism to do proofreading.
We perform this driving using a mechanical ratchet and pawl engine powered by a lowering weight.
In our pedagogical description of the model's operation, we implicitly assumed that this weight was very large
in order to enable the mechanism to proofread, similar to how the hydrolysis energy needs to be large for Hopfield's scheme to operate effectively.\cite{Hopfield1974}
In the full treatment of the model in section \ref{section:results}, however, we will demonstrate how the tuning
of the weight can give us graded levels of fidelity enhancement, and will also show that in the absence of
this weight the equilibrium fluctuations of the piston alone cannot lead to proofreading.}

In our model introduction we have also made 
several simplifying assumption for clarity of presentation which do not conform
with the principle of microscopic reversibility, and it is important that we relax them in the full
treatment of the model to make it thermodynamically consistent.
In particular, we assumed that the ligand is necessarily unbound and that the enzyme is necessarily inactive
when the piston is expanded, with the reverse assumptions made when the piston is compressed. 
We also assumed that substrate binding was prohibited in the active state of the enzyme.
These assumptions allowed us to claim that no premature product formation takes place in the expanded piston state
and that no unfiltered substrates bind to the activated enzyme in the compressed piston state, which, in turn, 
justified the use of long waiting times between the piston actions necessary to establish high levels of fidelity in
each piston state.
In the detailed analysis of our model presented in section \ref{section:results},
we relax these assumptions and 
consider the full diversity of enzyme states at each piston position with reversible transitions between them.
This, as we will demonstrate, 
will not only ensure the thermodynamic consistency of our treatment but will also
reveal the possibility of doing proofreading more than once
by leveraging the presence of multiple inactive intermediates in between enzyme's substrate-unbound
and production states which were not accounted for in our conceptual introduction of the model.
 
\newpage
\section{RESULTS}
\label{section:results}

{\color{black}
\subsection{Ratchet and Pawl Engine Enables a Tunable Control of Piston Actions}}
\label{section:engine}

To drive the cyclic compressions and expansions of the piston necessary for achieving proofreading,
we use a ratchet and pawl engine whose design is inspired by Feynman's original work. \cite{Feynman1963vol1}
In his celebrated lectures, Feynman presented two implementations of the ratchet and pawl engine -- 
one operating on the temperature difference between two thermal baths, 
and another driven by a weight that goes down due to gravity.
In the piston model we adopt the second scheme as it involves fewer parameters and 
illustrates the process of energy transduction more transparently.

\begin{figure*}[!ht]
	\centerline{
		\includegraphics[scale=1.00]{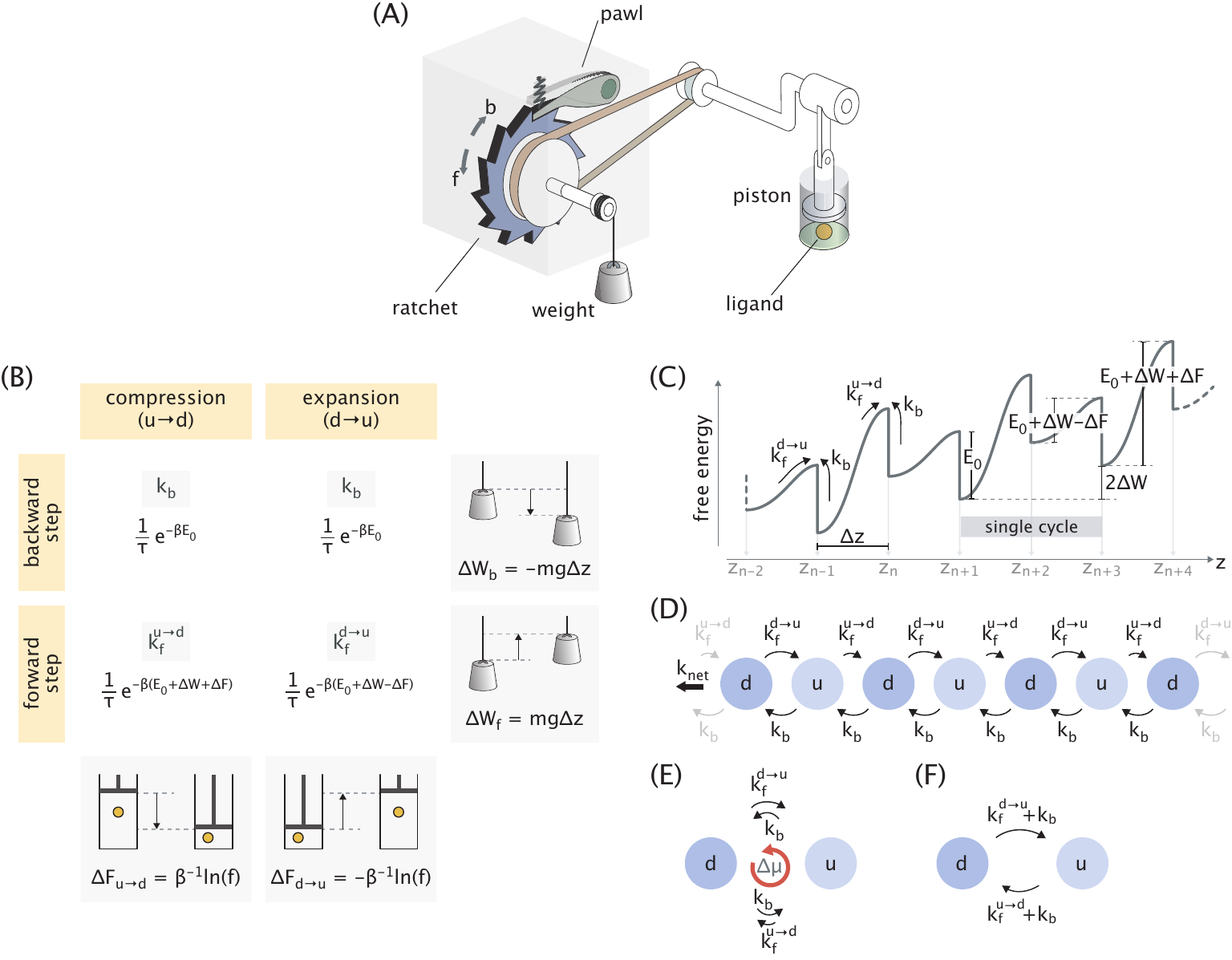}
	}
	\caption{Ratchet and pawl mechanism coupled to the piston. (A) Schematic representation of the mechanism.
	The different radii of the ratchet wheel and the axle of the crankshaft
	ensure that a single ratchet step translates into a full compression or a full expansion of the piston
	(i.e. a 180$^\circ$ rotation of the crankshaft).
	Arrows with symbols ``b'' and ``f'' indicate the directions of backward and forward ratchet rotation, respectively.
	(B) Rates of the four kinds of transitions (symbols in shaded boxes with explicit expressions below),
	along with the accompanying changes in the potential energy of the weight
	and the free energy of the ligand. (C) Free energy landscape corresponding to the non-equilibrium dynamics of the system
	in the presence of a non-zero weight. Discrete positions of the weight ($z_n$) 
	corresponding to energy minima of the landscape
	are marked on the reaction coordinate. (D) Infinite chain representation of the dynamics of discrete system states. 
	$k_\text{net}$ stands for the net rate at which the weight goes down.
	(E) Equivalent two-state representation of the engine dynamics where 
	the driving force $\Delta \mu = 2 \Delta W$ breaks the detailed balance in the diagram.
	(F) Collapsed representation of the diagram in panel E shown with the net transition rates from the two
	pathways.
	}
	\label{fig:engine_piston}
\end{figure*}

The ratchet and pawl engine coupled to the piston is shown in Figure~\ref{fig:engine_piston}A. 
The engine is powered by a weight of mass $m$ which is hanging from an axle connected to the ratchet.
The free rotational motion of the ratchet is rectified by a pawl; when the pawl sits on a ratchet tooth, it prevents
the ratchet from rotating in the clockwise (backwards) direction.
The mechanical coupling between the engine and the piston is achieved through a crankshaft mechanism
which translates each discrete ratchet step into a full compression ({\underline{u}p $\rightarrow$ \underline{d}own})
or a full expansion ({\underline{d}own $\rightarrow$ \underline{u}p}) of the piston.
We assume that the volume regulated by the piston contains a single ligand -- a choice motivated
by Szilard's thermodynamic interpretation of information, where a piston compressing a single gas
molecule was considered. \cite{Szilard1929}

The clockwise (backward) and counterclockwise (forward) steps of the microscopic ratchet are enabled
through environmental fluctuations.
Specifically, a backward step is taken whenever the pawl acquires sufficient energy from the environment
to lift itself over the ratchet tooth that it is sitting on, allowing the tooth to slip under it (hence, the name ``backward'').
Following Feynman's treatment,\cite{Feynman1963vol1} we write the rate of such steps as
\begin{align}
k_\text{b} = \tau^{-1} \text{e}^{-\beta E_0},
\end{align}
where $\tau^{-1}$ is the attempt frequency, $E_0$ is the amount of energy needed to lift the pawl over a ratchet tooth,
and $\beta = 1/k_\text{B} T$ is the inverse of the thermal energy scale 
(see Supporting Information section B.1
for a detailed discussion of the ratchet and pawl mechanism). 
Every backward step of the ratchet is accompanied by either a full compression  
or a full expansion of the piston,
as well as the lowering of the weight by an amount of $\Delta z$, which reduces its potential energy by 
$\Delta W = mg \Delta z$.

Unlike in backward stepping, for a forward step to take place, the rotational energy acquired by the ratchet through fluctuations
should be sufficient to not only overcome the resistance of the spring pressing the pawl onto the ratchet, 
but also to lift the weight
and to alter the state of the piston.
This is a pure consequence of the geometric design of the ratchet and the positioning of the pawl.
We assume that piston actions take place isothermally and in a quasistatic way, 
and therefore, write the changes in ligand free energy
upon compression (u$\, \rightarrow \,$d) and expansion (d$\, \rightarrow \,$u) as
\begin{align}
\dFud &= \beta^{-1} \ln f, \quad \text{and} \\
\dFdu &= - \beta^{-1} \ln f,
\end{align}
respectively, where $f = \Vup / \Vdown \ge 1$ is the fraction by which the volume occupied by the ligand decreases
upon compression.
The signs of free energy differences suggest that piston compressions slow down the forward steps, 
while expansions speed them up. 
These features are reflected in the two kinds of forward stepping rates that are given by
\begin{align}
\label{eqn:kfud}
\kfud &= \tau^{-1} \text{e}^{-\beta \left( E_0 + \Delta W + \Delta F \right)} = f^{-1} k_\text{b} e^{-\beta \Delta W}, \\
\label{eqn:kfdu}
\kfdu &= \tau^{-1} \text{e}^{-\beta \left( E_0 + \Delta W - \Delta F\right)} = f k_\text{b} e^{-\beta \Delta W},
\end{align}
where $\Delta F = \beta^{-1} \ln f$ and was used with a ``+'' and ``-'' sign in the place of
$\dFud$ and $\dFdu$, respectively. The rates of all four kinds of transitions, namely, forward or backward ratchet steps,
accompanied by either a compression or an expansion of the piston are summarized in Figure~\ref{fig:engine_piston}B.

In the presence of a nonzero weight ($\Delta W > 0$), the ratchet will on average rotate backwards -- a feature
reflected in the tilted free energy landscape shown in Figure~\ref{fig:engine_piston}C.
As can be seen, the average dissipation per step is $\Delta W$ and it is independent of $\Delta F$.
In addition, the work performed on the ligand upon compression is fully returned upon expansion,
which, as we will demonstrate in section \ref{section:tradeoffs},
will generally not be the case when we introduce the enzyme coupling.
To further study the nonequilibrium dynamics of the driving mechanism, we map the local minima of the energy landscape
corresponding to discrete vertical positions of the weight (equivalently, discrete ratchet angles) into an infinite chain
of transitions shown in Figure~\ref{fig:engine_piston}D. There ``d'' and ``u'' stand for the compressed and 
expanded states of the piston, respectively. 
The net stepping rate $\knet$ at which the weight goes down
can be written as
\begin{align}
\label{eqn:knet_2state}
\knet = \left( k_\text{b} - \kfdu \right) \pi_\text{d} + \left( k_\text{b} - \kfud \right) \pi_\text{u},
\end{align}
where $\pi_\text{d}$ and $\pi_\text{u}$ are the steady state probabilities of the compressed and expanded piston
states, respectively.
These probabilities can be obtained by considering the equivalent two-state diagram in Figure~\ref{fig:engine_piston}E
where the vertical position of the weight has been eliminated, and the nonequilibrium nature of the dynamics
is instead captured via the cycle through two alternative pathways connecting the piston states.
The driving force $\Delta \mu$ in this cycle is given by\cite{Hill2012}
\begin{align}
\label{eqn:chem_potential}
\Delta \mu = \beta^{-1} \ln \left( \frac{k^2_\text{b}}{\kfdu \kfud} \right) = 2 \Delta W,
\end{align}
demonstrating the broken detailed balance in the presence of a nonzero weight, and confirming
the dissipation of $2\Delta W$ per cycle observed in the energy landscape (Figure~\ref{fig:engine_piston}C).
We note that this procedure of mapping a linear network onto a cyclic one has also been used 
for modeling the processivity of molecular motors, where the linear coordinate corresponds to the 
position of the motor while the alternating states correspond to different motor conformations.\cite{Qian1997, WagonerDill2016}

\begin{figure*}[!ht]
	\centerline{
		\includegraphics[scale=0.94]{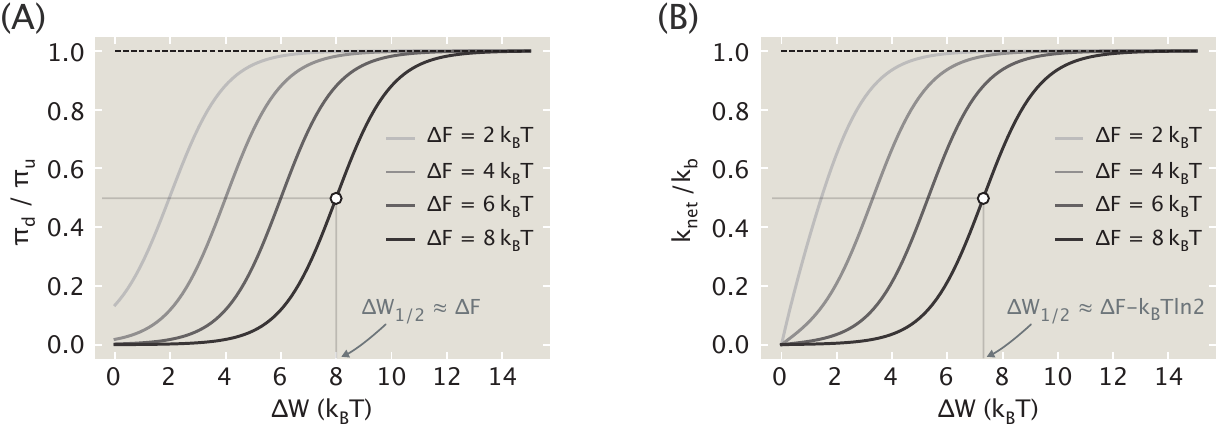}
	}
	\caption{Nonequilibrium features of the engine-piston coupling.
	(A) Steady state probability ratio of compressed (``d'') and expanded (``u'') piston states 
	and (B) normalized net rate of backward stepping ($k_\text{net}/k_\text{b}$) as a function of
	the work per step ($\Delta W$) for different choices of the ligand compression energy ($\Delta F$).
	The $\Delta W_{1/2}$ expressions stand for the values of $\Delta W$ 
	where the corresponding value on the $y$-axis is 0.5
	(Supporting Information section B.2).
	Negative $\Delta W$ values are not considered as they further increase the undesired bias in 
	piston state occupancies.
	}
	\label{fig:piston_tuning}
\end{figure*}

At steady state, the net incoming and outgoing fluxes at each piston state 
in Figure~\ref{fig:engine_piston}E should cancel each other (seen more vividly in the collapsed
diagram in Figure~\ref{fig:engine_piston}F), namely,
\begin{align}
\label{eqn:steady_2state}
\left( \kfdu + k_\text{b} \right) \pi_\text{d} = \left( \kfud + k_\text{b} \right) \pi_\text{u}.
\end{align}
Substituting the expressions for forward stepping rates (eqs~\ref{eqn:kfud} and \ref{eqn:kfdu}) into eq~\ref{eqn:steady_2state}
and additionally imposing the probability normalization constraint ($\pi_\text{d} + \pi_\text{u} = 1$), 
we can solve for $\pi_\text{d}$ and $\pi_\text{u}$ to obtain
\begin{align}
\label{eqn:pid}
\pi_\text{d} &= \frac{1 + \text{e}^{-\beta(\Delta W + \Delta F)}}{2(1 + \cosh (\beta \Delta F) \, \text{e}^{-\beta \Delta W} )}, \\
\label{eqn:piu}
\pi_\text{u} &= \frac{1 + \text{e}^{-\beta (\Delta W - \Delta F)} } {2(1 + \cosh (\beta \Delta F) \, \text{e}^{-\beta \Delta W} )}.
\end{align}
Notably, in the absence of external drive ($\Delta W = 0$), piston state occupancies follow the Boltzmann 
distribution, that is, $(\pi_\text{d}/\pi_\text{u})_\text{eq} = \text{e}^{-\beta \Delta F} = f^{-1}$, suggesting that 
at equilibrium the piston will predominantly dwell in the expanded state.
Conversely, as can be seen in Figure~\ref{fig:piston_tuning}A, when the work per step exceeds $\Delta F$ by several $k_\text{B} T$, the occupancies of the two piston states become equal to each other.
This happens because at large $\Delta W$ values forward ratchet stepping becomes very unlikely and the dynamics
proceeds only through backward steps with a rate $k_\text{b}$ which is identical for both compressive and expansive steps.
As will be shown in section \ref{section:tradeoffs}, 
suppressing this equilibrium bias set by $\Delta F$ is essential
for achieving efficient proofreading, analogous to the need for driving the transitions between the two enzyme-substrate intermediates in Hopfield's scheme (Figure~\ref{fig:piston_model_concept}A). \cite{Hopfield1974}

With the steady state probabilities known, we can now substitute them into eq~\ref{eqn:knet_2state} 
to find the net rate at which the weight goes down, obtaining
\begin{align}
\label{eqn:knet}
k_\text{net} = \frac{\left( 1 - \text{e}^{-2\beta \Delta W}\right) k_\text{b}}{1 + \cosh(\Delta F) \text{e}^{-\beta \Delta W}}.
\end{align}
As expected, $k_\text{net}$ vanishes at equilibrium ($\Delta W = 0$), and asymptotes to $k_\text{b}$ 
at large $\Delta W$ values, as shown in Figure~\ref{fig:piston_tuning}B.
The knowledge of $k_\text{net}$ allows us to calculate the power ($P$) dissipated for the maintenance of the
nonequilibrium steady state. Specifically, since $k_\text{net}$ is the rate at which the weight goes down and
$\Delta W$ is the dissipation per step, the power $P$ becomes their product, namely,
\begin{align}
P &= k_\text{net} \Delta W.
\end{align}

The formalism developed in this section for characterizing the steady state behavior of the system will be used as
a basis for defining the different performance metrics of the model in section \ref{section:coupling}.

\newpage
\subsection{Thermodynamic Constraints Make Fidelity Enhancement Unattainable in the Absence of External Driving}
\label{section:enzyme}

In order to implement a thermodynamically consistent coupling between the engine and the allosteric enzyme,
we need to consider the full diversity of possible enzyme states,\cite{EinavJPCB2016} and not just the dominant ones 
depicted in Figure~\ref{fig:piston_model_concept}B.
Therefore, in this section we provide a comprehensive discussion of the enzyme in 
an equilibrium setting before introducing its coupling to the engine.

\begin{figure*}[!ht]
	\centerline{
		\includegraphics[scale=1.00]{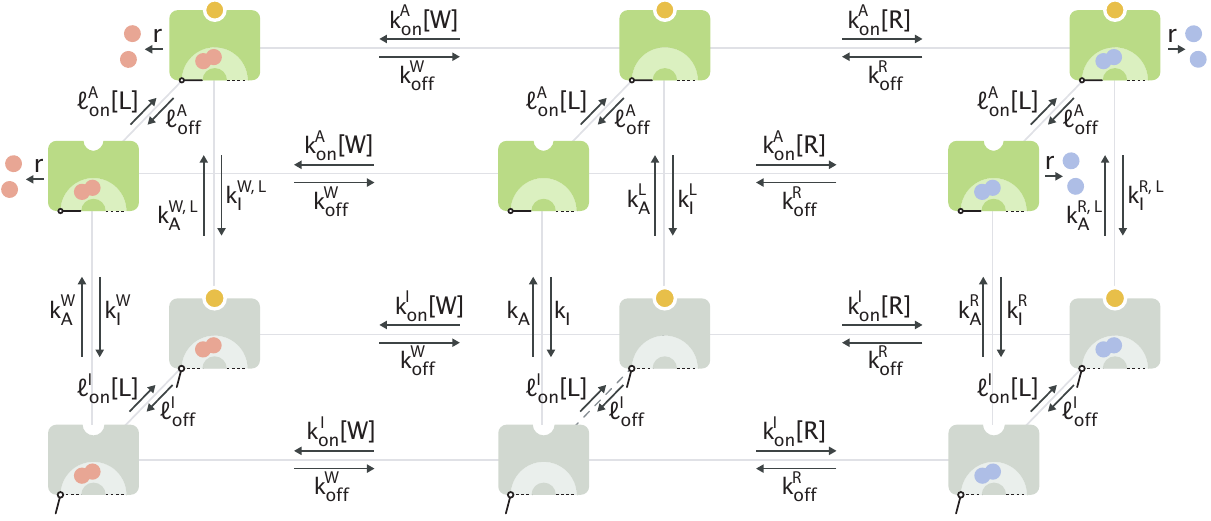}
	}
	\caption{
	Network diagram of enzyme states and transitions between them.
	Right (``R'') and wrong (``W'') substrates are depicted in blue and red, respectively.
	The orange circle represents the ligand (``L''). 
	Active (``A'') and inactive (``I'') enzymes are shown in green and gray, respectively.
	}
	\label{fig:enzyme_equil}
\end{figure*}

The network diagram of all possible enzyme states is depicted in Figure~\ref{fig:enzyme_equil}.
As can be seen, each of the twelve states are defined by enzyme's catalytic activity and
whether or not a ligand and a right/ wrong substrate are bound to the enzyme.
Following the principle of microscopic reversibility, \cite{Tolman1979}
we assign non-zero rate constants to the transitions between enzyme states.
Only the product formation (with rate $r$) is taken to be 
an irreversible reaction under the assumption that the system is open
where the formed products are taken out and an influx of new substrates is maintained.
Since in our model neither the enzyme nor the substrates carry an energy source, 
the choice of the different rate constants cannot be arbitrary.
Specifically, the cycle condition needs to be satisfied for each closed loop of the diagram,
requiring the product of rate constants in the clockwise direction to equal the product
in the counterclockwise direction (Supporting Information section C.1).\cite{Hill2012}

With these equilibrium restrictions imposed on the rate constants, 
we can show that when the ligand concentration is held fixed ($[\text{L}](t)$ = const)
the fidelity of the enzyme cannot exceed that defined by the 
ratio of the off-rates, namely, $\koffW/\koffR$
(see Supporting Information section C.2).
What allows the enzyme to beat this equilibrium limit of fidelity
without direct coupling to hydrolysis is the
cyclic alteration of the ligand concentration between low and high values (thus, $[\text{L}](t) \ne$ const).
In our model, we achieve this cyclic alteration through the ratchet and pawl engine driving the piston actions - 
a choice motivated by our objective to provide an explicit treatment of energy management.
We note, however, that fidelity enhancement can be achieved irrespective of the driving agency 
as long as the cyclic alteration of ligand concentration is maintained at 
a certain ``resonance'' frequency, 
the presence of which we demonstrate in section~\ref{section:tradeoffs}.

\newpage
\subsection{Coupling the Engine to the Enzyme Gives the Full Description of the Piston Model}
\label{section:coupling}

\begin{figure*}
	\centerline{
		\includegraphics[scale=1.00]{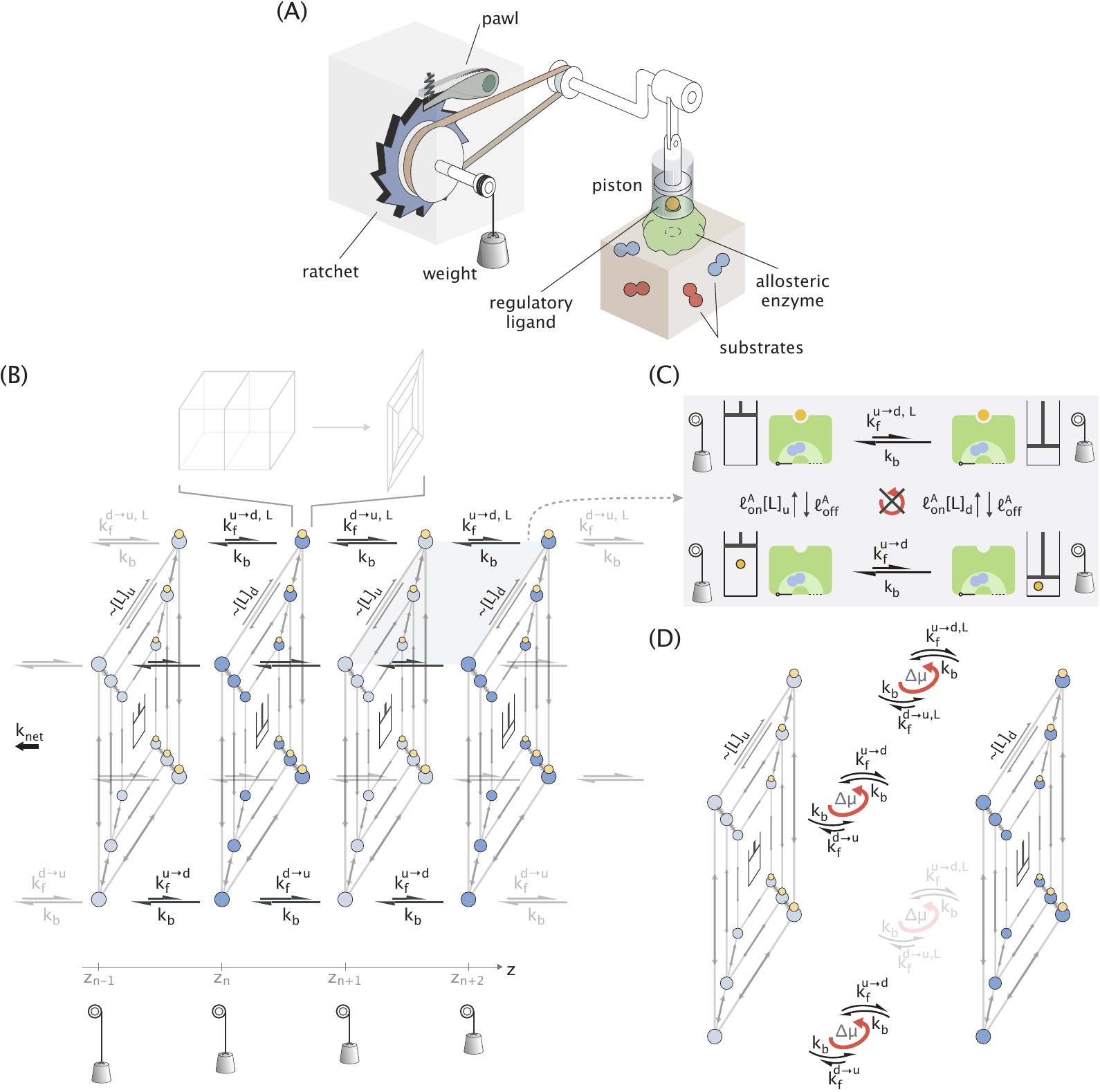}
	}
	\caption{Full and thermodynamically consistent treatment of the piston model of proofreading.
	(A) Schematic representation of the full model, with the ratchet and pawl engine coupled to the enzyme.
	(B) Network diagram of the full model. Each slice of the diagram represents the planar view of the enzyme state network,
	with the alternating colors corresponding to the compressed (dark blue) and expanded (light blue) states of the piston.
	Ligand-bound enzyme states are marked with an orange circle.
	The horizontal arrows connecting the slices stand 
	for forward and backward ratchet steps. Only those at the outer edges are shown
	for clarity; however, transitions are present between all horizontally neighboring enzyme states.
	Also for clarity, the stepping rate constants are shown only at two of the outer edges where the ligand is either unbound
	(bottom edge) or bound to the enzyme (top edge).
	The hanging weight at different vertical positions is displayed below the diagram to symbolize energy expenditure as it gets
	lowered with a net rate $\knet$.
	(C) Sub-network of the full diagram in panel B where the state of the system is unchanged after doing a
	cyclic traversal.
	The red arrow with a cross on top indicates that the cycle condition holds in the sub-network.	
	(D) The finite-state equivalent of the full network in panel B with the weight position ($z_n$) eliminated. 
	Red arrows indicate the driving with a force $\Delta \mu = 2 \Delta W$.
	}
	\label{fig:full_coupling}
\end{figure*}

Having separately introduced the driving mechanism in section \ref{section:engine} and 
the allosteric enzyme with the full diversity of its states in section \ref{section:enzyme}, 
we now couple the two together to obtain the full driven version of the piston
model, shown in Figure \ref{fig:full_coupling}A.
The coupling is achieved by exposing the ligand binding site of the enzyme to the piston compartment 
where the activator ligand is present.
The enzyme can therefore ``sense'' the state of the piston (and, thereby, the effects of driving) through the induced 
periodic changes in the ligand concentration.

In the absence of enzyme coupling, the network diagram capturing the nonequilibrium dynamics of the system was 
an infinite one-dimensional chain (Figure~\ref{fig:engine_piston}D), where each discrete state was defined by the vertical
position of the weight ($z_n$) and the state of the piston (``u'' or ``d'').
In the layout where the engine and the enzyme are coupled, the full specification of the system state now
requires three items: the position of the weight ($z_n$), the piston state (``u'' or ``d''),
and the state of the enzyme (one of the 12 possibilities).
By converting the three-dimensional view of the enzyme state network (Figure~\ref{fig:enzyme_equil}) into 
its planar equivalent, we represent the nonequilibrium dynamics of this coupled layout
again through an infinite chain, but this time each 
slice at a fixed weight position ($z_n$)
corresponding to the planar view of the enzyme state network (Figure \ref{fig:full_coupling}B).
The slices alternate between the compressed and expanded piston states (dark and light blue circles, respectively),
with high and low ligand concentrations used in the transition network inside each slice.

Arrows between the slices (not all of them shown for clarity) represent the forward and backward steps of the ratchet.
Crucially, as a consequence of coupling, the rates of forward stepping now depend on the state of the enzyme.
In particular, when the ligand is bound to the enzyme, it no longer exerts pressure on the piston and therefore,
in those cases, the forward stepping rates become simply
\begin{align}
\label{eqn:kfudL}
\kfudL = \kfduL = \kb \text{e}^{-\beta \Delta W},
\end{align}
where the superscript ``L'' indicates that the ligand is bound (orange circles in Figure~\ref{fig:full_coupling}B).
We note that in the general case with $N$ ligands, the pressure would drop down to that of $(N-1)$ ligands upon ligand binding,
correspondingly altering the rates of forward stepping
(see Supporting Information section D.1 for details).
This adjustment of forward rates is essential for the thermodynamic consistency of coupling the engine to the enzyme.
Specifically, it ensures that any cycle of transitions that brings the enzyme and the weight back into their original states
is not accompanied by dissipation,
consistent with the fact that in the piston model energy is spent only when there is a net lowering of the weight.
As a demonstration of this feature, consider the cycle in Figure~\ref{fig:full_coupling}C which is extracted from the
larger network. 
Using the expression of forward stepping rates in eqs~\ref{eqn:kfud} and \ref{eqn:kfudL}, we can write the cycle condition
for this sub-network as
\begin{align}
\frac{\kfudL \times \loffA \times \kb  \times \lonA [\text{L}]_\text{u}}{\kb \times \loffA \times \kfud \times \lonA [\text{L}]_\text{d}} = 
\frac{[\text{L}]_\text{u} f}{[\text{L}]_\text{d}} = 1,
\end{align}
where the equality $f = \Vup/\Vdown = [\text{L}]_\text{d}/[\text{L}]_\text{u}$ was used.
The fact that the products of rate constants in clockwise and counterclockwise directions are identical shows that
no dissipation occurs when traversing the cycle.

Now, to study how driving affects the proofreading performance of the piston model, we need to obtain
the steady state probabilities of the different enzyme states.
To that end, we convert the full network diagram into an equivalent form shown in 
Figure \ref{fig:full_coupling}D, where 
we have eliminated the position of the weight ($z_n$), akin to the earlier treatment of the uncoupled
engine in Figure~\ref{fig:engine_piston}E.
Note that the transitions between the two slices again 
represent piston compression and expansion events driven by 
a force $\Delta \mu = 2\Delta W$, as in eq~\ref{eqn:chem_potential}.
The steady state probabilities $\pi_i$ of the 24 different states in Figure~\ref{fig:full_coupling}D 
(12 enzyme states $\times$ 2 piston states) can be obtained from the set of all rate constants,
the details of which we discuss in Supporting Information section D.2.
With these probabilities known, 
we calculate the rate of energy dissipation ($P$), 
speed of forming right products ($\vR$), and fidelity ($\eta$) as
\begin{align}
\label{eqn:power_full}
P &= \underbrace{\sum_{i = 1}^{24} \left( \kb - \kf^{(i)} \right) \pi_i}_{\knet} \times \Delta W, \\
\vR &= \sum_{i \in S^{\mathrm{A}}_{\mathrm{R}}} \pi_i \times r, \\
\eta &= \frac{\vR}{\vW} = \frac{\sum_{i \in S^{\mathrm{A}}_{\mathrm{R}}} \pi_i }{\sum_{i \in S^{\mathrm{A}}_{\mathrm{W}}} \pi_i},
\end{align}
where $\kf^{(i)}$ is the rate constant of making a forward step from the $i^{\mathrm{th}}$ state ($1 \le i \le 24$),
while $S_{\mathrm{R}}^{\mathrm{A}}$ and $S_{\mathrm{W}}^{\mathrm{A}}$ are the sets of catalytically active 
enzyme states with a right and wrong substrate bound, respectively.

One significant downside of using these ``raw'' 
metrics in the numerical evaluation of the model performance is their high sensitivity
to the particular choices of parameter values. 
We therefore introduce their scaled alternatives which we will use for the numerical studies in sections 
\ref{section:tradeoffs} and \ref{section:enzyme_tuning}.
Specifically, as a measure of energetic efficiency, we use the dissipation per right product formed,
defined as
\begin{align}
\varepsilon = \frac{P}{\vR}.
\end{align}
This way, the metric of energetics has units of $\kB T$ and is independent
from the choice of absolute timescale.
Then, as a dimensionless metric of speed, we introduce the normalized quantity
\begin{align}
\nu = \frac{\vR}{v_\text{R}^\text{MM}},
\end{align}
which represents 
the fraction by which the rate of forming right products in the proofreading setting ($\vR$) 
is slower than that in the simple Michaelis-Menten scheme ($v_\text{R}^\text{MM}$)
where the allosteric effects are absent. This normalizing Michaelis-Menten
speed is given in terms of the model parameters via 
\begin{align}
v_\text{R}^\text{MM} = \frac{\frac{\konI [\text{R}]}{\koffR + r}}{1 + \frac{\konI [\text{R}]}{\koffR + r} + \frac{\konI [\text{W}]}{\koffW + r}} \times r.
\end{align}
Next, we define the proofreading index $\alpha$ as a fidelity metric which represents the degree to which
the fidelity is amplified in multiples of $\koffW/\koffR$ over its Michaelis-Menten value ($\eta_\text{MM}$),
that is,
\begin{align}
\eta &= \underbrace{\left( \frac{\koffW+r}{\koffR + r} \right)}_{\eta_\text{MM}} \left( \frac{\koffW}{\koffR} \right)^\alpha, \\
\alpha &= \frac{\ln \eta - \ln \eta_\text{MM}}{ \ln \left( \frac{\koffW}{\koffR} \right)}.
\end{align}
Note that the proofreading index of Hopfield's scheme is $\alpha_\text{Hopfield} = 1$, as it involves a single
proofreading realization.
Also, since in the absence of external driving the highest fidelity is 
$\eta_\text{eq}^\text{max} = \koffW/\koffR$, the corresponding upper limit in the proofreading index becomes
$\alpha_\text{eq} = 1 - \ln \eta_\text{MM} / \ln \eta_\text{eq}^\text{max}$.

As a final descriptor of piston model's nonequilibrium behavior, we introduce the fraction
of returned work ($\kappa$) defined as the ratio of the rate at which 
the ligand performs work on the piston upon expansion
to the rate at which the piston performs work on the ligand upon compression. 
We calculate $\kappa$ via
\begin{align}
\kappa = -\frac{\sum_{i \in S_\text{d}} \left(\kb + \kf^{(i)} \right) \pi_i \Delta F_{\text{d}\rightarrow\text{u}}^{(i)}}
{\sum_{i \in S_\text{u}} \left(\kb + \kf^{(i)} \right) \pi_i \Delta F_{\text{u}\rightarrow\text{d}}^{(i)}},
\end{align}
where $S_\text{d}$ and $S_\text{u}$ are the sets of states where the piston is compressed and expanded,
respectively. The negative sign is introduced to account for the fact that the ligand free energy decreases
upon piston expansion (i.e. the system gets the work back).
In the absence of enzyme coupling (section \ref{section:engine}), this ratio was 1 because the ligand
constantly exerted pressure on the piston.
With enzyme coupling, however, the work performed on the ligand upon compression may not be fully returned
since with some probability the ligand will be bound to the enzyme and exert no pressure on the piston during expansion.
We therefore expect $\kappa$ to be generally less than 1, indicating a net rate of performing work on the ligand in the
nonequilibrium setting.

Having defined analytical expressions for the key model performance metrics,
we now proceed to studying their graded changes and the trade-offs between them numerically.

\newpage
\subsection{Energy-Speed-Fidelity Trade-Off in the Piston Model}
\label{section:tradeoffs}

Because of its mechanical construction, 
the piston model of proofreading has a distinguishing feature - 
in it the external driving mechanism is physically separated from the allosteric enzyme.
This feature allows us to independently examine how tuning the ``knobs'' of the engine
and varying the kinetic parameters of the enzyme alter the performance of the model.

\begin{figure*}
	\centerline{
		\includegraphics[scale=1.00]{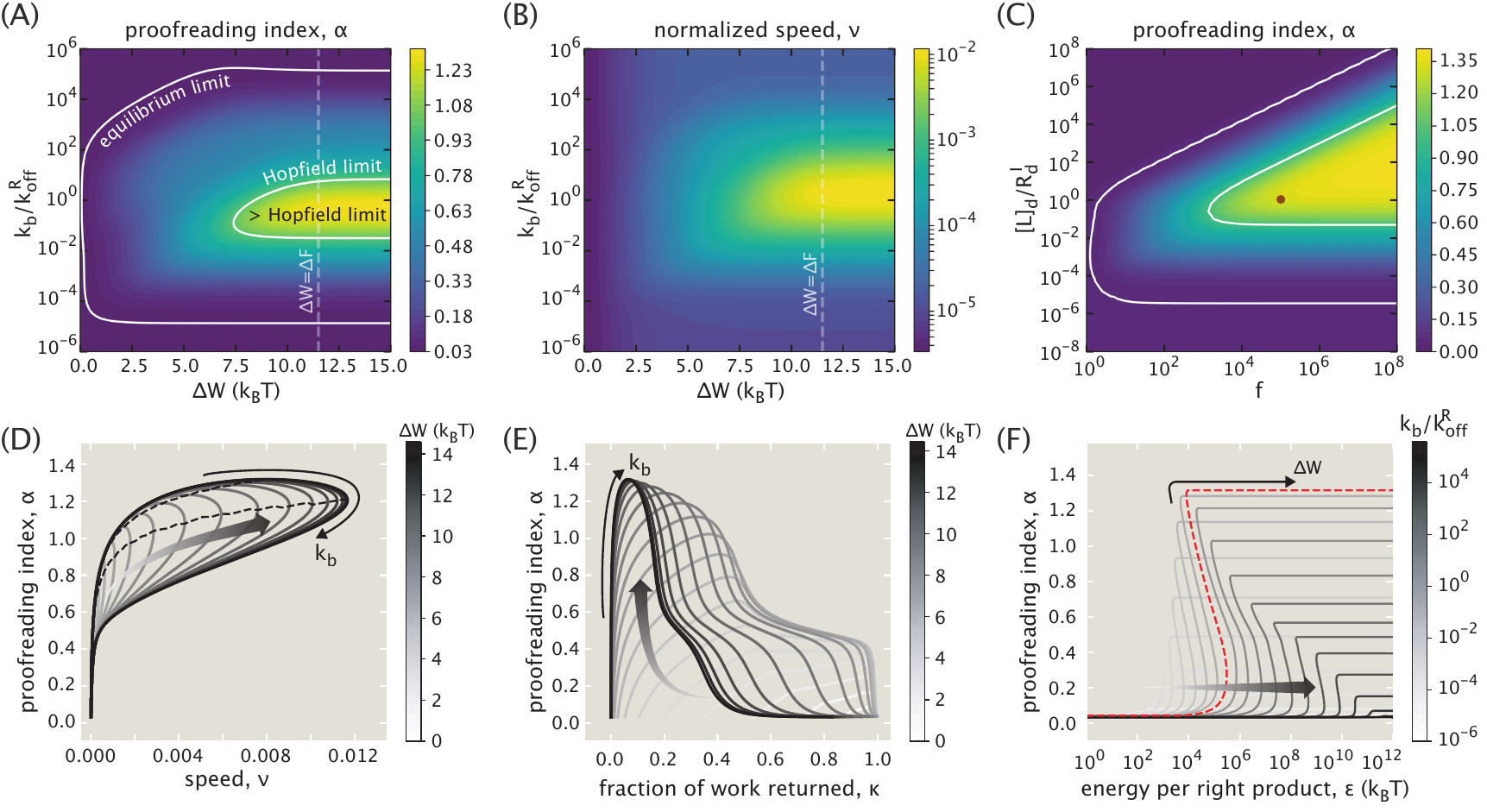}
	}
	\caption{Parametric studies on the changes in the piston model performance in response to
	tuning the ``knobs'' of the engine. 
	(A,B) Variations in the proofreading index (A) and speed (B) as the rate of backward stepping ($\kb$)
	and the work per step ($\Delta W$) are tuned.
	The dotted line corresponds to the value of $\Delta W$ equal to the ligand free energy change upon
	compression ($\Delta F$).
	(C) Variations in the proofreading index when the high ligand concentration ($[\text{L}]_\text{d}$) 
	and the compression factor ($f$) are tuned.
	$R_\text{d}^\text{I}$ represents the ligand dissociation constant in the inactive enzyme state.
	The red dot indicates the pair of $[L]_\text{d}$ and $f$ values used in the studies of the other panels.
	(D) Fidelity-speed trade-off as $\kb$ is continuously varied for different choices of $\Delta W$ 
	(gradient arrow shows the direction of increase).
	The dotted black lines connect the highest fidelity and speed values as $\Delta W$ is tuned.
	Between these dotted lines fidelity and speed are negatively correlated.
	(E) Relation between fidelity and fraction of returned work for discrete choices of $\Delta W$
	and continuously tuned $\kb$ values
	(the gradient arrow indicates increasing $\Delta W$).
	(F) Fidelity-dissipation trade-off obtained by continuously tuning $\Delta W$ for discrete choices
	of the hopping rate ($\kb$). The gradient arrow indicates the direction of increasing $\kb$.
	{\color{black}The red dotted curve corresponds to the case with resonance $\kb$.}
	}
	\label{fig:trade_offs}
\end{figure*}

We begin our numerical analysis by first exploring the effects of external driving, where the tuning ``knobs''
include the rate of backward stepping ($\kb$), the work per step ($\Delta W$), 
the ligand concentration in the compressed piston state ($[\text{L}]_\text{d}$)
and the compression factor ($f = [\text{L}]_\text{d}/[\text{L}]_\text{u}$).
Choosing a set of enzyme's kinetic parameters which make proofreading possible
(see Supporting Information section D.3
for the full list of parameters), we keep them fixed for the rest of the analysis.
We conduct the first parametric study by tuning $\kb$ and $\Delta W$
and evaluating the proofreading index (Figures~\ref{fig:trade_offs}A).
As anticipated, the proofreading index does not exceed its equilibrium limit
in the absence of driving ($\Delta W = 0$). 
This expected feature can be paralleled by Brownian motors 
where purely equilibrium fluctuations of the motor's energy landscape are unable to generate directed motion. \cite{AstumianHanggi2002}
In addition, the proofreading index achieves its highest value
if $\Delta W$ is comparable to or larger than the ligand compression energy $\Delta F$,
and if the backward hopping rate $\kb$ is at its ``resonance'' value.
The presence of a ``resonance'' hopping rate is intuitive since if piston actions take place
very slowly then the 
fidelity will be reduced due to the small but nonzero rate of forming unfiltered products (i.e. ``leakiness'')
in the quasi-equilibrated enzyme states.
And, conversely, if 
piston actions take place too rapidly, then the activator ligand will almost always be bound to the enzyme,
preventing the realization of multiple substrate discrimination layers through sequential enzyme activation
and inactivation.
We note that analogous resonance responses were also identified for Brownian particles which
attain their highest nonequilibrium drift velocity in a ratchet-like potential landscape when 
the temperature\cite{Reimann1996} or the landscape profile\cite{Faucheux1995} are temporally
modulated at specific resonance frequencies.
A similar feature is present in Hopfield's model as well; namely, optimal proofreading is attained only when the rate of
hydrolysis is neither too low, nor too high.\cite{WongPRE2018}
Interestingly, when the driving is hard enough ($\Delta W \gtrsim \Delta F$) and the backward hopping rate is close to its resonance value,
the fidelity of the piston model 
beats the Hopfield limit ($\alpha=1$) and raises the question of the largest attainable fidelity,
which we discuss in the next section.

Trends similar to those for the proofreading index are observed for the speed of forming right products as well
(Figure~\ref{fig:trade_offs}B).
Specifically, product formation is very slow in the absence of driving and increases monotonically with $\Delta W$,
until plateauing when $\Delta W \gtrsim \Delta F$.
Also, the highest speed is achieved at a resonance $\kb$ value which is different from that of the proofreading index.
The existence of such a resonance frequency is again intuitive, since at fast rates of piston actions 
the enzyme is predominantly active and unable to bind new substrates, while at slow rates 
the enzyme activation for catalysis via piston compression happens very rarely.
Notably, since the enzyme parameters were chosen in a way so as to yield high fidelities, the largest speed value
is substantially lower than the corresponding speed for a single-step Michaelis-Menten enzyme 
($\nu_\text{max} \approx 10^{-2}$).

In the last parametric study, we explore how the choice of the high and low ligand concentrations affects the 
performance of the model.
To that end, we tune the high ligand concentration ($[\text{L}]_\text{d}$) and 
the compression factor ($f = [\text{L}]_\text{d}/[\text{L}]_\text{u}$), 
and evaluate the highest proofreading index at the resonant $\kb$ value with $\Delta W > \Delta F$.
As we can see, large fidelity enhancements are achieved when $[\text{L}]_\text{d}$ is comparable to or larger than
the ligand dissociation constant in the inactive enzyme state ($R_\text{d}^\text{I}$), 
which is necessary to activate the enzyme upon piston
compression. In addition, the compression factor needs to be large enough (or, equivalently, the ligand concentration
in the expanded piston state should be low enough) so as to inactivate the enzyme when piston enters its expanded state.
This requirement of having a large free energy difference between the compressed and expanded
piston states ($\beta \Delta F = \ln(f) \gg 1$) 
can be paralleled with a similar condition in Hopfield's model where for optimal proofreading
the energy of the activated enzyme-substrate complex needs to be much larger than that of the inactive complex.

Knowing separately how tuning the engine ``knobs'' affects the fidelity and speed, we now explore the trade-offs
between the model's performance metrics as we vary the driving parameters $\kb$ and $\Delta W$, while holding the
high and low ligand concentrations at fixed values (the red dot in Figure~\ref{fig:trade_offs}C).
We start with the trade-off between fidelity and speed, depicted in Figure~\ref{fig:trade_offs}D, where we 
continuously tune the hopping rate $\kb$ for different choices of the driving force $\Delta W$.
As expected from the results of the individual parametric studies in Figure~\ref{fig:trade_offs}A and B, 
both fidelity and speed increase monotonically with $\Delta W$. 
Also, since the values of the hopping rate $\kb$ that maximize fidelity and speed are not identical,
these two performance metrics are negatively correlated in the range of $\kb$ values defined by the two 
different resonance rates (region between the dotted lines in Figure~\ref{fig:trade_offs}D), 
but are positively correlated otherwise.
Variations in the metrics in the region of their negative correlation, however, are moderate, suggesting that
for an allosteric enzyme that has been optimized for doing proofreading, the largest speed and fidelity
could be achieved at similar external driving conditions.

Next, we consider the relation between fidelity and fraction of work returned, shown in Figure \ref{fig:trade_offs}E.
As can be seen, no fidelity enhancement is achieved when $\kappa$ is close to $1$ which happens
either in the absence of driving (lighter curves) or in the presence of driving, provided that the hopping
rate is very fast. 
On the other hand, $\kappa$ is much less than $1$ at the peak fidelity which is achieved when the hopping rate
is at its resonance value and when driving is large ($\Delta W \gtrsim \Delta F$).
Overall, this trade-off study demonstrates that irreversible work performed on the ligand is
a required feature for the attainment of fidelity enhancement in the piston model.

Lastly, we look at how fidelity varies with energy dissipation, with the latter characterized through the energy
expended per right product ($\varepsilon$). The results of the trade-off study are shown in Figure~\ref{fig:trade_offs}F.
There, the driving force is continuously tuned for different choices of the hopping rate.
As can be seen, there is a minimum dissipation per product required to attain the given
level of performance. This minimum dissipation (the first intercept at a given y-level) is achieved when 
the hopping rates are less than the corresponding resonant values (the lighter curves on the left side
of the dotted red curve).
Additionally, for a given hopping rate, increasing the driving force ($\Delta W$) 
could lead to an increased proofreading performance and a
decreased dissipation per product up a critical point where the performance metric reaches its saturating value
(horizontal region),
demonstrating how increasing the driving force could in fact improve the energetic efficiency of proofreading.
{\color{black}
We note here that the minimum $\varepsilon$ values needed for significant proofreading are $\sim 10^3 - 10^4 \, \kB T$
in Figure~\ref{fig:trade_offs}F which is $\sim 2$ orders of magnitude higher than what is 
calculated for translation by the ribosome.\cite{WongPRE2018}
This low energetic efficiency can be a consequence of 
our particular parameter choice for the study as well as the performance limitations of our engine design,
the investigation of which we leave to future work.}

\newpage
\subsection{Up to Three Proofreading Realizations are Available to the Piston Model}
\label{section:enzyme_tuning}

In the previous section, we chose a set of kinetic rate constants for the enzyme and, keeping them fixed,
explored the effects of tuning the external driving conditions on the performance of the model.
In this section, we explore the parameter space from a different angle, namely, we study how tuning the
enzyme's kinetic parameters changes the model performance under optimal driving conditions.
Since there are more than a dozen rates defining the kinetic behavior of the enzyme, 
it is impractical to 
probe their individual effects.
Instead, we choose to vary two representative parameters about the effects of which we have a prejudice.
These include the rate of substrate binding to the active enzyme ($\konA$) and the unbinding rate of 
wrong substrates ($\koffW$).
We know already from Hopfield's analysis that for efficient proofreading the direct binding of substrates to 
the active enzyme state should be 
very slow. 
Therefore, we expect the proofreading performance to improve
as $\konA$ is reduced. We also expect the minimum requirement for $\konA$ to be lower 
for larger $\koffW$ values to ensure that wrong substrates do not enter through the unfiltered pathway. \cite{Hopfield1974}

\begin{figure*}
	\centerline{
		\includegraphics[scale=1.00]{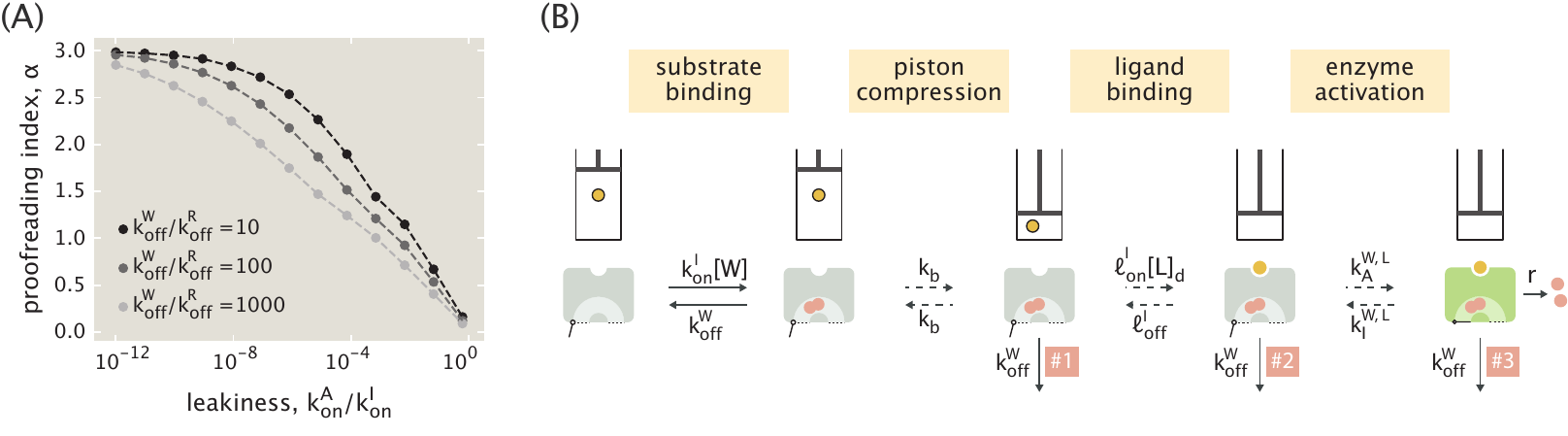}
	}
	\caption{
	Proofreading performance of the piston model under optimized enzyme parameters and external
	driving conditions.
	(A) The highest proofreading index ($\alpha$) available to the piston model as a function of leakiness 
	($\konA/\konI$) for different choices of $\koffW$.
	{\color{black}
	(B) The dominant trajectory that the system takes to form a wrong product in the case where $\konA/\konI = 10^{-12}$.
	Numbers 1, 2, 3 stand for the different proofreading filters along the trajectory.
	The dotted arrows indicate that the respective rates are much slower than the substrate unbinding rate $\koffW$ (see Supporting Information section D.5 for their numerical values for the $\koffW/\koffR = 100$ case).}
	}
	\label{fig:multi_proof}
\end{figure*}

With these expectations in mind, we performed a parametric study to find the highest fidelity, the results of which
are summarized in Figure~\ref{fig:multi_proof}A.
There we varied $\konA$ for several choices of $\koffW$, and for each pair numerically optimized 
over the enzyme's
remaining kinetic rates and external driving conditions to get maximum fidelity (see
Supporting Information section D.4 for implementation details).
As expected, the highest attainable fidelity 
decreases monotonically with increasing ``leakiness'' ($\konA/\konI$),
and the minimum requirement on $\konA$ decreases with increasing $\koffW$.

Interestingly, we also see that for small enough leakiness, the piston model manages to perform proofreading 
(i.e. enhance the fidelity by a factor of $\koffW/\koffR$)
up to three times, as $\alpha_\text{max} \approx 3$ (Figure~\ref{fig:multi_proof}A).
{\color{black}
To understand this unexpected feature, we identified the dominant trajectory that the system would take to form a
wrong product for the case where $\konA/\konI = 10^{-12}$ 
(Figure~\ref{fig:multi_proof}B, see Supporting Information section D.5 for details).
As we can see, after initial binding the wrong substrate indeed passes through three different proofreading
filters, and these are realized efficiently because 
the transitions between intermediate states are much slower than the rate of substrate unbinding.
The first filter occurs right after piston compression, 
while the enzyme is waiting for the activator ligand to bind (\#1).
We note that this particular filter is made possible due to the presence of alternative piston states
(equivalently, alternative environments that the enzyme could ``sense'').
The remaining two filters (\#2 and \#3) take place while the ligand-bound enzyme is waiting to get activated
and while the active enzyme is waiting to turn the wrong substrate into a product, respectively.
The presence of these two filters is purely a consequence of allostery.
Importantly, the $\alpha_\text{max} \approx 3$ result in Figure~\ref{fig:multi_proof} represents 
the theoretical upper limit of the model's proofreading index -- a feature that we justify analytically in 
Supporting Information section D.5.

In light of this analysis, we can now explain why the pedagogically simplified version of the model introduced in
section~\ref{section:model} achieved only a single proofreading realization. There we made the
implicit assumption that 
ligand binding after piston compression and enzyme activation after ligand binding took place instantly. 
Because of this, proofreading filters \#1 and \#2 were not realized, leaving filter \#3 as the only available one
which we showed in Figure~\ref{fig:piston_model_concept}B.}

\newpage
\section{Discussion and Conclusions}

A distinctive feature of kinetic proofreading is that it is a nonequilibrium mechanism, by virtue of which its operation
needs to involve energy expenditure.\cite{Hopfield1974,NinioBiochimie1975}
Mechanical work, being an intuitive representation of energy expenditure, has been used in the past to elucidate important
physical concepts such as information-to-energy conversion in the thought experiment by Szilard,\cite{Szilard1929}
or the mechanical equivalence of heat in Joule's apparatus.\cite{Joule1850}
Yet, a similar demonstration of how mechanical work could be harnessed in a graded fashion
to beat the equilibrium limit in substrate discrimination fidelity has been lacking.
Our aim in this work was to offer such a demonstration through the mechanically designed piston model of proofreading.

We started off by providing the conceptual picture of the piston model, with its constituents having direct parallels with
Hopfield's original proofreading scheme\cite{Hopfield1974} (Figure~\ref{fig:piston_model_concept}).
The key idea of the model was to replace the nucleotide hydrolysis step present in Hopfield's scheme with 
piston compression which served an identical role of activating the enzyme but in our case achieved through allostery
and mechanical work.
Just like in the case of biological proofreading, 
where hydrolysis itself cannot lead to fidelity enhancement unless 
the nucleotide triphosphates are held at fixed out-of-equilibrium chemical potentials,
in the case of piston model too, 
the compressive and expansive actions of the piston cannot result in proofreading unless
they are driven by an energy-consuming engine.
Motivated by Feynman's ratchet and pawl mechanism, \cite{Feynman1963vol1}
we then proposed a dissipative mechanical engine to drive 
the cyclic piston actions,
which maintained the nonequilibrium distribution of enzyme states necessary for achieving proofreading.
The function of this engine can be paralleled to that of the ATP synthase in the cell whose constant operation maintains 
a finite chemical potential of ATP, which different biochemical pathways can then take advantage of.

To study how the cyclic variations in ligand concentration generated by the engine alter the occupancies of enzyme states,
we performed a thermodynamically consistent coupling between the engine and the enzyme
(Figure~\ref{fig:full_coupling}).
There we considered the full diversity of states that the enzyme could take and, importantly,
the feedback mechanism for the engine to ``sense'' the state of the enzyme.
The accounting of this latter feature, which makes the piston model an example of a bi-partite system,
\cite{HorowitzEsposito2014, BaratoNewJPhys2014}
was motivated by our aim of proposing a framework where 
we could consistently calculate
the total dissipation 
as opposed to only the minimum dissipation needed for maintaining the nonequilibrium steady state of the enzyme
(without considering the driving engine). \cite{WongPRE2018, CuiMehta2018, HorowitzPRE2017}
{\color{black}
Although the dissection of different contributions to dissipation 
and their interconnectedness was not among the objectives of our work,
the framework that we proposed in our model can serve as a basis for additional studies of periodically driven
molecular systems (e.g. Brownian clocks or artificial molecular motors) 
where the driving protocol and thermodynamics are of importance.
\cite{SeifertNewJPhys2017, BaratoClock2016, LeighNature2003}}
As noted earlier, however, in the presence of a periodically changing ligand concentration, the allosteric enzyme
could perform proofreading irrespective of the driving agency, which suggests a possible biochemical mechanism of fidelity enhancement
without the direct coupling of the enzyme state transitions to hydrolysis.

Having explicit control over the ``knobs'' of the mechanical engine, 
we then probed the performance of the model under different driving conditions.
We found that both speed and fidelity increased as we tuned up the mass of the hanging weight, until plateauing 
at a point where the free energy bias of the expanded piston state was fully overcome ($\Delta W \gtrsim \Delta F$),
beyond which increasing the weight only increased the dissipation without improving the model performance
(Figure~\ref{fig:trade_offs}A,B).
This result can be paralleled with the presence of a minimum threshold for the strength of driving in Hopfield's model,
past which the highest fidelity becomes attainable.\cite{Hopfield1974}
In addition, we found that in the piston model there is a ``resonance'' rate of piston actions which maximizes fidelity,
analogous to 
the similar feature of Hopfield's scheme where both very fast and very slow rates of hydrolysis reduce the
quality of proofreading.\cite{WongPRE2018}

The tunable control over the driving parameters also allowed us to study the trade-off between 
fidelity, speed and energy spent per right product.
These studies revealed that the correlation between speed and fidelity could be both positive and negative
when varying the rate of driving.
Notably, theoretical investigations of translation by the \textit{Escherichia coli} ribosome under Hopfield's scheme
identified a similar behavior for the fidelity-speed correlation in response to tuning the GTP hydrolysis rate,
with the experimentally measured values being in the negative correlation (i.e. trade-off) region.\cite{BanerjeePNAS2017}
In contrast to the ribosome study, however, 
where the two metrics vary by several orders of magnitude in the trade-off region,
in the piston model the variations in fidelity and speed in the negative correlation region are 
moderate (Figure~\ref{fig:trade_offs}D),
calling for additional investigations of the underlying reasons behind this difference and search for the realization of the
latter advantageous behavior in biological proofreading systems.
Furthermore, our studies showed that the minimum dissipation required to reach the given
level of fidelity was achieved for hopping rates necessarily lower than their resonance values, and that 
increasing the work performed per step (analogously, the chemical potential of ATP) 
could actually improve the energetic efficiency of the model -- features that again motivate the identification
of their realization in biochemical systems.

In the end, we explored the limits in the proofreading performance of the piston model 
for various choices of the allosteric enzyme's ``leakiness'' ($\konA/\konI$) 
and the ratio of the wrong and right substrate off-rates ($\koffW/\koffR$).
We found that the trends for the highest available fidelity matched analogously
with the features of Hopfield's original scheme,
suggesting their possible ubiquity for general proofreading networks.
More importantly, our analysis revealed that the piston model could do proofreading not just once 
but up to three times in the limit of very low leakiness, 
despite the fact that energy consumption takes place during a single piston compression.
This is in contrast to the typical involvement of several energy consumption instances
in multistep proofreading schemes which manage to beat the Hopfield limit of fidelity,
as, for example, in the cases of the T-cell or MAPK activation pathways which require multiple phosphorylation reactions.
\cite{Mckeithan1995, CuiMehta2018, SwainBiophysJ2002}
Our finding therefore suggests the possibility of achieving several proofreading realizations with a single energy consuming
step by leveraging the presence of multiple inactive intermediates intrinsically available to allosteric molecules.
We would like to mention here that the presence of a similar feature was also 
experimentally demonstrated recently for the ribosome which was shown to use the free energy of a single GTP hydrolysis
to perform proofreading twice after the initial tRNA selection -- first, at the $\text{EF-Tu}\cdot\text{GDP}$-bound inactive state and second, at the EF-Tu-free active state.\cite{IeongPNAS2016}

In the presentation of the piston model we focused on the thermodynamic consistency of the framework
for managing the energy dissipation and did not consider strategies for improving the performance of the mechanism.
One such possibility that can be considered in future work is to use a more elaborate design 
for the ratchet and pawl engine with alternating activation barriers for pawl hopping 
which would allow to have different rates of piston compression and expansion,
analogous to how hydrolysis and condensation reactions generally occur with different rates in biological proofreading.
\cite{BanerjeePNAS2017, ChenNelson2018}
Another avenue is to consider alternative ways of allocating 
the mechanical energy dissipation across the different ratchet transition steps,
similar to how optimization schemes of allocating the free energy of ATP hydrolysis were studied
for molecular machine cycles.\cite{BrownSivakPNAS2017}
Incorporating these additional features would allow us to probe the performance limits of the piston model
and compare them with the fundamental limits set by thermodynamics.\cite{Qian2006}

\section*{Acknowledgements}
We thank Tal Einav, Erwin Frey, Christina Hueschen, Sarah Marzen, Arvind Murugan, 
Manuel Razo-Mejia, Matt Thomson, Yuhai Tu, Jin Wang, Jerry Wang, 
Ned Wingreen and Fangzhou Xiao for fruitful discussions.
We also thank Haojie Li and Dennis Yatunin for their input on this work, 
Alexander Grosberg, 
David Sivak, and Pablo Sartori for providing valuable feedback on the manuscript,
and Nigel Orme for his assistance in making the illustrations.
This work was supported by 
the National Institutes of Health through the grant 1R35 GM118043-01 (MIRA),
and the John Templeton Foundation 
as part of the Boundaries of Life Initiative grants 51250 and 60973.

\newpage
\bibliography{PaperLibrary}

\end{document}


\maketitle
\singlespacing

 \appendix
 
\setcounter{equation}{0}
\setcounter{figure}{0}
\setcounter{page}{1}
\renewcommand{\thepage}{S\arabic{page}}
\renewcommand{\thefigure}{S\arabic{figure}}
\renewcommand{\thetable}{S\arabic{table}}
\renewcommand{\theequation}{S\arabic{equation}}

\renewcommand\contentsname{Contents}
\tableofcontents
\addtocontents{toc}{\protect\setcounter{tocdepth}{2}}

\newpage
\section{Discrimination Fidelity in the Conceptual Scheme of the Piston Model}
\label{section:fidelity_concept}
In this section, we derive the expressions for the fidelities achieved at the two piston steps introduced in 
section 2 of the main text, namely, $\eta_1 = \koffW/\koffR$ when the piston is expanded, 
and $\eta_2 = (\koffW + r)/(\koffR+r)$ when the piston is compressed. In our discussion, we retain the 
simplifying assumptions made during the presentation of the model concept.

\begin{figure*}[!ht]
	\centerline{
		\includegraphics[scale=1.00]{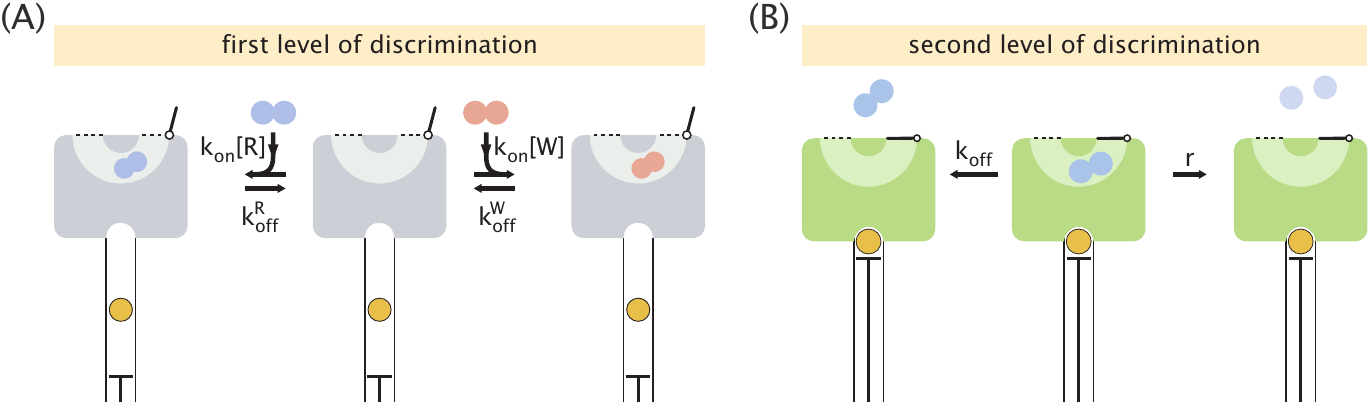}
	}
	\caption{The two substrate discrimination levels in the conceptual scheme of the piston model.
	(A) The first level is achieved when the piston is expanded and a roughly equilibrium distribution
	of substrate-bound and free enzyme states is established.
	(B) The second level is achieved in the compressed state of the piston where the enzyme is active 
	and can either release the bound substrate or turn it into a product.
	}
	\label{fig:fidelities_conceptual}
\end{figure*}

As discussed in section 2, the first level of substrate discrimination occurs in the expanded piston state.
If the waiting time for compression is long enough for the substrates to equilibrate with the inactive enzyme,
we can impose the detailed balance condition at the two pairs of edges in Figure \ref{fig:fidelities_conceptual}A to obtain
\begin{align}
\label{eqn:si_detailed_r}
\pR_\text{I} \koffR &= p_\text{I} \, k_\text{on} [\text{R}], \\
\label{eqn:si_detailed_w}
\pW_\text{I} \koffW &= p_\text{I} \, k_\text{on} [\text{W}].
\end{align}
Here $p_\text{I}$, $\pR_\text{I}$ and $\pW_\text{I}$ stand for the probabilities of the empty, right substrate-bound and
wrong substrate-bound inactive states of the enzyme, respectively. 
Taking the substrate concentrations to be identical ($[\text{R}] = [\text{W}]$),
we can equate the left sides of eqs~\ref{eqn:si_detailed_r} and \ref{eqn:si_detailed_w} to find
\begin{align}
\frac{\pR_\text{I}}{\pW_\text{I}} = \frac{\koffW}{\koffR}.
\end{align}
The above ratio of probabilities represents the proportion in which right and wrong substrate-bound inactive enzymes
enter the active state, and therefore, becomes equivalent to the fidelity $\eta_1$ achieved in the first discrimination step.

The second level of substrate differentiation takes place when the piston gets compressed, leading to the activation
of the enzyme. We assume that in its active state
the enzyme can no longer bind new substrates. If we wait long enough,
a substrate that was already bound before piston compression
will either unbind with a rate $k_\text{off}$ or get turned into a product with a rate $r$.
The probability that a product is formed can be written as
\begin{align}
\label{eqn:prod_prob}
p_\text{prod} = \int_{0}^\infty \mathrm{d}t' \, p_\text{bound} (t') \times r,
\end{align}
where $p_\text{bound} (t')$ is the probability that the substrate is still bound by time $t'$.
Using the fact that the waiting time distribution of substrate release 
(either through unbinding or product formation)
is $P_\text{release}(t) = (k_\text{off} + r) e^{-(k_\text{off} + r)t}$, the probability $p_\text{bound} (t)$ can be found as
\begin{align}
p_\text{bound}(t) = \int_{t}^\infty \mathrm{d}t' \, P_\text{release}(t') = e^{-(k_\text{off} + r)t}.
\end{align}
Substituting this result into eq~\ref{eqn:prod_prob} and performing the integration, we obtain
\begin{align}
p_\text{prod} = \frac{r}{k_\text{off} + r}.
\end{align}
Due to the difference in the off-rates of the right and wrong substrates, their respective probabilities of production will also
be different, resulting in the second level of fidelity given by the ratio of these probabilities, namely,
\begin{align}
\eta_2 = \frac{p_\text{prod}^\text{R}}{p_\text{prod}^\text{W}} = \frac{\koffW + r}{\koffR + r}.
\end{align}

\newpage
\section{Ratchet and Pawl Engine}

In this section, we first provide a detailed discussion of the ratchet and pawl mechanism in the absence of piston
coupling.
Then, for the case of piston coupling, we derive of the expressions for the work per step ($\Delta W_{1/2}$) 
shown in Figure 3 at which the 
ratio of piston state probabilities ($\pid/\piu$) and the net rate of backward stepping ($\knet$)
reach 50\% of their respective saturation values.

\subsection{Details of the Ratchet and Pawl Mechanism in the Absence of Piston Coupling}
\label{section:ratchet_pawl_engine_details}
\begin{figure*}[!ht]
	\centerline{
		\includegraphics[scale=1.00]{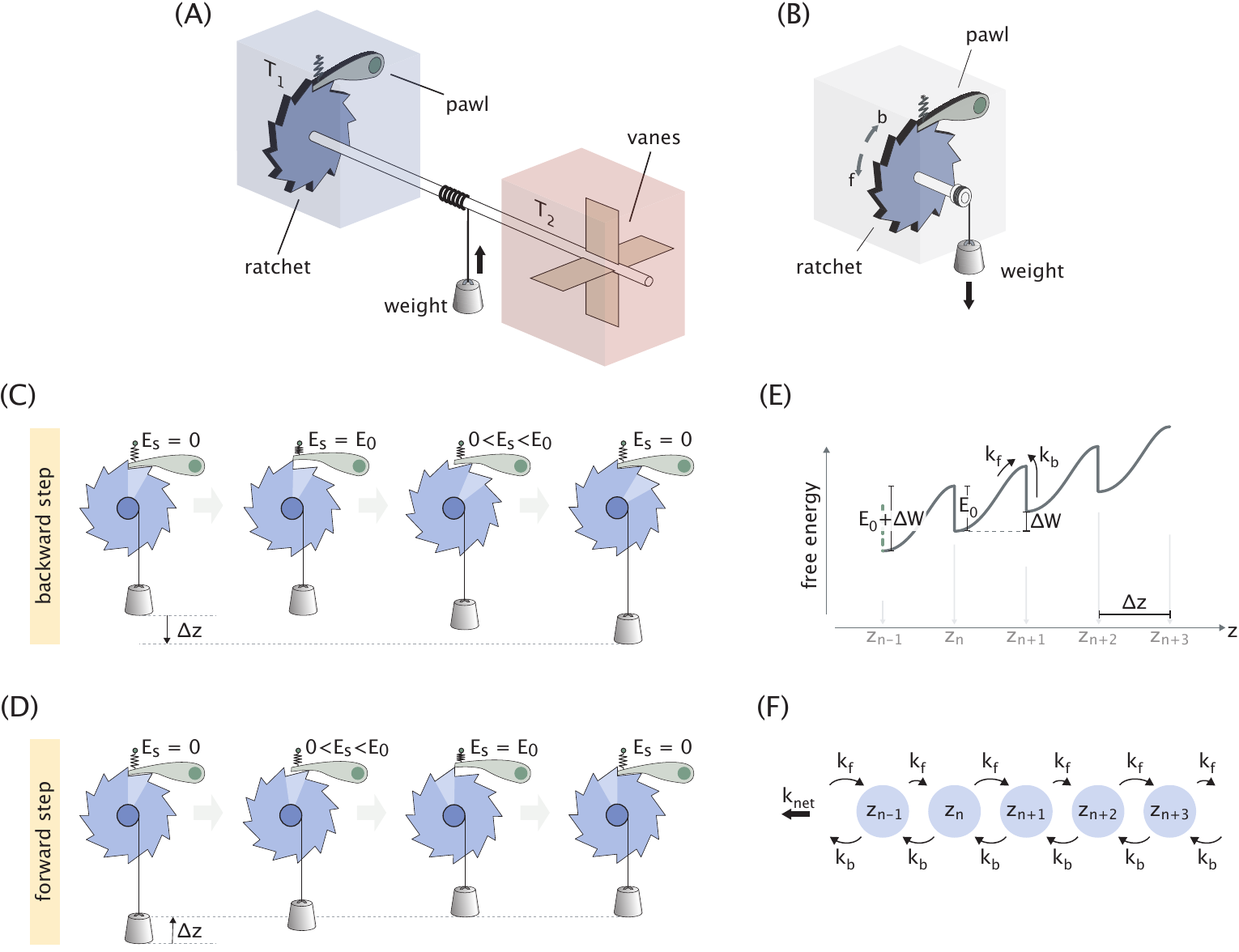}
	}
	\caption{Working details of the ratchet and pawl mechanism.
	(A) Feynman's original ratchet and pawl mechanism operating as a heat engine. \cite{Feynman1963vol1}
	(B) Ratchet and pawl engine driven by a hanging weight that is used in the piston model.
	Arrows with symbols ``b'' and ``f'' indicate the directions of backward and forward ratchet rotation, respectively.
	(C)-(D) Breakdown of backward (C) and forward (D) steps of the ratchet, accompanied by the lowering 
	or the lifting of the weight, respectively. $E_\text{s}$ stands for the potential energy of the spring.
	(E) Free energy landscape corresponding to the directionally biased rotations of the ratchet due 
	to a net lowering of the weight. Discrete positions of the weight ($z_n$) corresponding to the energy minima of the 
	landscape are marked on the reaction coordinate.
	(F) Infinite chain representation of the discrete state dynamics. When a non-zero weight is hung from the axle,
	the ratchet makes backward steps with a net rate $\knet$.}
	\label{fig:ratchet_pawl_basics}
\end{figure*}

The ratchet and pawl mechanism was originally proposed by Richard Feynman with an aim to demonstrate
the validity of the second law of thermodynamics.\cite{Feynman1963vol1}
In his description, the mechanism had an additional element, namely, vanes that were connected to the ratchet
through a massless axle (Figure~\ref{fig:ratchet_pawl_basics}A).
The purpose of the vanes was to induce forward ratchet steps through thermal
fluctuations.
When the temperature in the vane compartment was maintained at a higher value than that in the ratchet compartment
($T_2 > T_1$),
the mechanism could utilize this difference to operate as a heat engine and lift a weight hanging from the axle.

In the piston model, instead of running the ratchet and pawl mechanism as a heat engine, we drive it
at a constant temperature
through the expenditure of the gravitational potential 
energy of the hanging weight.
We have therefore removed the vane compartment from our description of the engine and 
ascribed forward stepping to random 
rotational fluctuations of the ratchet instead (Figure~\ref{fig:ratchet_pawl_basics}B).

As mentioned in section 3.1, backward stepping takes place whenever the pawl borrows sufficient energy from the
environment
to overcome the potential energy barrier $E_0$ of the spring and lift itself over the ratchet tooth that it is
sitting on, allowing the tooth to slip under it (Figure~\ref{fig:ratchet_pawl_basics}C).
Once the pawl gets over the ratchet tooth (step 2 in Figure~\ref{fig:ratchet_pawl_basics}C), 
the hanging weight and the recovering pawl start applying torque on the ratchet, causing it to rotate in the 
clockwise direction (step 3 in Figure~\ref{fig:ratchet_pawl_basics}C).
Following Feynman's treatment, 
we assume that when the pawl hits the bottom of the next tooth (step 4 in Figure~\ref{fig:ratchet_pawl_basics}C),
the total kinetic energy of the system, which is the sum of the energy borrowed by the pawl
and the change in the potential
energy of the weight per step ($\Delta W = mg \Delta z$), 
gets dissipated due to the perfectly inelastic collision of the pawl
with the ratchet.
Therefore, as a result of a single backward step, the net heat dissipated into the environment becomes 
$\Delta W$, as reflected in the free energy landscape in Figure~\ref{fig:ratchet_pawl_basics}E.

A similar set of arguments for forward stepping would imply that initially the mechanism needs to borrow enough
energy from the environment to overcome the spring barrier and to lift the weight by an amount of $\Delta z$ 
(step 3 in Figure~\ref{fig:ratchet_pawl_basics}D). We again assume, that once the pawl passes over the next tooth
and inelastically hits the ratchet, it dissipates all its accumulated potential energy. Therefore, in the end of a single 
forward step, the total energy extracted from the environment is equal to the increase in the potential energy of the weight per forward step (Figure~\ref{fig:ratchet_pawl_basics}E).

Implicit in our treatment of ratchet stepping has been the assumption that we could discretize the possible configurations
of the mechanism into states where the pawl fully rests on ratchet teeth. Within this formalism, we took $E_0$ to be
the activation energy of backward stepping and $(E_0 + \Delta W)$ to be the activation energy of forward stepping, 
resulting in rate constants given by
\begin{align}
\kb &= \tau^{-1} \text{e}^{-\beta E_0}, \\
\kf &= \tau^{-1} \text{e}^{-\beta (E_0 + \Delta W)},
\end{align}
where $\tau^{-1}$ is the attempt frequency. The choice of identical attempt frequencies for forward and backward steps
is, in a way, a requirement in the discretization formalism
to ensure that in the absence of driving ($\Delta W = 0$) no net rotation of the ratchet is generated, since
$\knet = \kb - \kf$ (Figure~\ref{fig:ratchet_pawl_basics}F). 
We note that a more rigorous treatment of ratchet stepping kinetics would need to account for the precise
shape of the energy landscape, defined both by the position of the weight (equivalently, the ratchet angle)
and the angular position of the fluctuating pawl, similar to the analysis done
by Magnasco and Stolovitzky.\cite{Magnasco1998}

\subsection{Derivation of $\Delta W_{1/2}$ Expressions}
\label{section:dW50_derivations}
We begin by deriving the $\Delta W_{1/2}$ expression at which the ratio $\pid/\piu$ is $1/2$.
From eqs 12 and 13, this ratio, evaluated at $\Delta W_{1/2}$, can be written as
\begin{align}
\frac{\pid}{\piu} \bigg|_{\Delta W_{1/2}} = \frac{1 + \text{e}^{-\beta(\Delta W_{1/2} + \Delta F)}}{1 + \text{e}^{-\beta(\Delta W_{1/2} - \Delta F)}} = \frac{1}{2}.
\end{align}
Solving for $\Delta W_{1/2}$, we obtain
\begin{align}
\Delta W_{1/2} &= \Delta F + \ln \left( 1 + \text{e}^{-2\beta \Delta F}\right) \nonumber\\
&= \Delta F + \ln \left( 1 +  f^{-2} \right),
\end{align}
where in the last step we used the expression for the ligand free energy change written 
in terms of the compression factor, that is, $\Delta F = \beta^{-1} \ln f$. Since for efficient
proofreading the compression factor needs to be large ($f \gg 1$), the $\Delta W_{1/2}$ expression
reduces into
\begin{align}
\Delta W_{1/2} \approx \Delta F.
\end{align}
To estimate how much the work per step needs to exceed $\Delta F$ in order for the ratio $\pid/\piu$
to reach its saturation value of $1$, we calculate the derivative of the ratio at $\Delta W_{1/2} \approx \Delta F$, namely,
\begin{align}
\frac{\partial}{\partial \Delta W} \left( \frac{\pid}{\piu} \right) \bigg|_{\Delta W_{1/2}} &= 
\frac{-\beta \, \text{e}^{-\beta(\Delta W_{1/2} + \Delta F)} \left( 1 + \text{e}^{-\beta(\Delta W_{1/2} - \Delta F)} \right) +  \left( 1 + \text{e}^{-\beta(\Delta W_{1/2} + \Delta F)} \right)  \, \beta \text{e}^{-\beta (\Delta W_{1/2} - \Delta F)} }{(1 + \text{e}^{-\beta (\Delta W_{1/2} - \Delta F)})^2} \nonumber\\
&= \frac{\beta(1 - \text{e}^{-2\beta \Delta F})}{4} \nonumber\\
&\approx \frac{1}{4 k_\text{B} T},
\end{align}
where we again employed the $\text{e}^{-2\beta \Delta F} \ll 1$ approximation. These results indicate that
in order to overcome the equilibrium bias in piston state probabilities caused by the higher ligand entropy
in the expanded state, 
the work per step needs to exceed the ligand free energy change upon compression ($\Delta F$) by
several $k_\text{B} T$ values.

Now, we perform a similar set of calculations for the net rate of backward stepping ($\knet$).
Using its expression in eq 14, we obtain
\begin{align}
\knet \big|_{\Delta W_{1/2}} = 
\frac{\left( 1 - \text{e}^{-2\beta \Delta W_{1/2}}\right) k_\text{b}}{1 + \cosh(\beta \Delta F) \text{e}^{-\beta \Delta W_{1/2}}} = \frac{\kb}{2}
\end{align}
Rearranging the terms, we obtain a quadratic equation for $\text{e}^{\beta \Delta W_{1/2}}$, namely,
\begin{align}
\text{e}^{2\beta \Delta W_{1/2}}  - \cosh (\beta \Delta F) \text{e}^{\beta \Delta W_{1/2}}  - 2 = 0.
\end{align}
Since $\text{e}^{\beta \Delta W_{1/2}} > 0$, we take the positive solution and obtain
\begin{align}
\text{e}^{\beta \Delta W_{1/2}} = \frac{\cosh(\beta \Delta F) + \sqrt{ \cosh^2(\beta \Delta F) + 8}}{2}.
\end{align}
For large degrees of compression ($\text{e}^{\beta \Delta F} \gg 1$), we can make the approximation 
$\cosh(\beta \Delta F) \approx \text{e}^{\beta \Delta F}/2$ and ignore the constant term in the square root,
which yields
\begin{align}
\label{eqn:w50df}
\text{e}^{\beta \Delta W_{1/2}} &\approx \frac{\text{e}^{\beta \Delta F}}{2}, \\
\Delta W_{1/2} &\approx \Delta F - \beta^{-1} \ln 2.
\end{align}
Like in the treatment of the ratio $\pid/\piu$, we now estimate how much the work per step needs to 
exceed $\Delta W_{1/2}$ in order for the backward stepping rate ($\knet$) to reach its saturating value $\kb$.
To that end, we calculate the derivative of $\knet/\kb$ at $\Delta W_{1/2}$, namely,
\begin{align}
\frac{\partial}{\partial \Delta W} \left( \frac{\knet}{\kb} \right) \bigg|_{\Delta W_{1/2}} &= 
\frac{ 2\beta \text{e}^{-2\beta \Delta W_{1/2}} (1 + \cosh(\beta \Delta F) \text{e}^{-\beta \Delta W_{1/2}}) + 
(1 - \text{e}^{-2\beta \Delta W_{1/2}}) \cosh(\Delta F) \beta \text{e}^{-\beta \Delta W_{1/2}}
}{ \left( 1 + \cosh(\beta \Delta F) \text{e}^{-\beta \Delta W_{1/2}} \right)^2} \nonumber\\
&\approx \frac{
2\beta \text{e}^{-2\beta \Delta W_{1/2}} (1 + \frac{1}{2} \text{e}^{\beta \Delta F} \text{e}^{-\beta \Delta W_{1/2}})
+ (1 - \text{e}^{-2\beta \Delta W_{1/2}}) \frac{1}{2} \text{e}^{\beta \Delta F} \beta \text{e}^{-\beta \Delta W_{1/2}}
}{(1 + \frac{1}{2} \text{e}^{\beta \Delta F} \text{e}^{- \beta \Delta W_{1/2}})^2} \nonumber\\
&= \frac{4\beta \text{e}^{-2\beta \Delta W_{1/2}} + \beta (1 - \text{e}^{-2\beta \Delta W_{1/2}})}{4} \nonumber\\
&\approx \frac{1}{4 k_\text{B} T}.
\end{align}
Here we made the approximation $\cosh(\Delta F) \approx \text{e}^{\beta \Delta F}/2$ in the first step, 
used the result from
eq~\ref{eqn:w50df} to write $\frac{1}{2} \text{e}^{\beta \Delta F} \text{e}^{- \beta \Delta W_{1/2}} = 1$ in the
second step,
and in the last step ignored the $\text{e}^{- 2 \beta \Delta W_{1/2}}$ terms since from eq~\ref{eqn:w50df} 
we have $\text{e}^{- 2 \beta \Delta W_{1/2}} = 4 \text{e}^{-2 \beta \Delta F} \ll 1$ for large degrees of compression.

As we can see, when the work per step exceeds $\Delta F$ by several $k_\text{B} T$ values, 
the chances of forward stepping become vanishingly small compared with backward stepping,
resulting in a net backward stepping rate $\knet \approx \kb$.

\newpage
\section{Equilibrium Properties of the Allosteric Enzyme}
In this section, we introduce the constraints on the choices of rate constants for the enzyme stemming from
the cycle condition (based on the fact that is does not consume energy), and also, discuss the fidelity available
to it when the ligand concentration is held at a fixed value.

\subsection{Constraints on the Choice of Enzyme's Rate Constants}
\label{section:enzyme_cycle}
\begin{figure*}[!ht]
	\centerline{
		\includegraphics[scale=1.00]{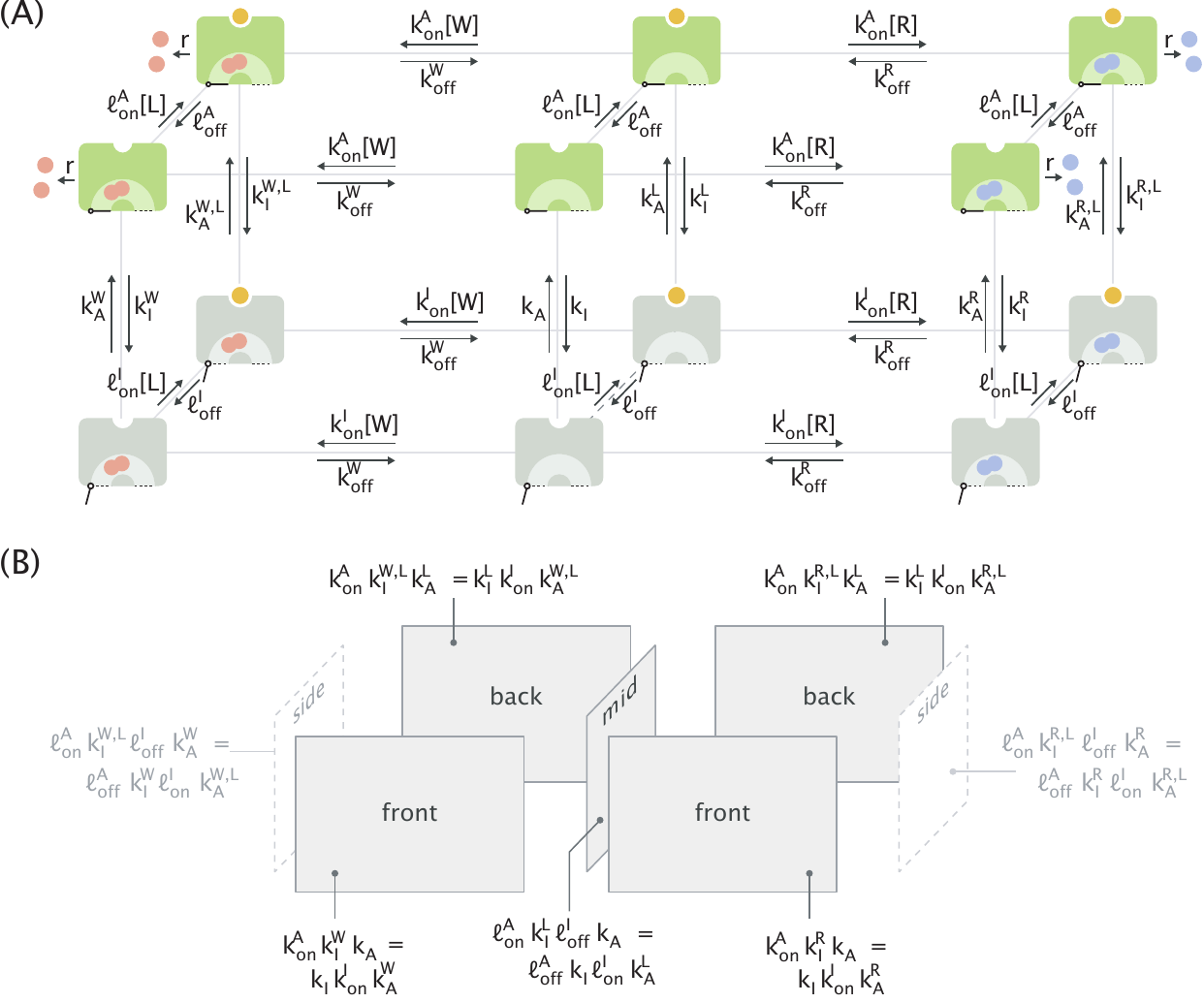}
	}
	\caption{The allosteric enzyme in the absence of engine coupling.
	(A) Network diagram of enzyme states and transitions between them at a fixed ligand concentration.
	The diagram is redrawn identically from Figure 4 for convenience.
	(B) Cycle condition on rate constants applied for the different loops of the diagram.
	The lighter color of the side loop conditions
	indicates that they are redundant and follow from the conditions on the other three loops.}
	\label{fig:enzyme_equilibrium_SI}
\end{figure*}

Consider the network of enzyme states and transitions in the absence of engine coupling redrawn in Figure~\ref{fig:enzyme_equilibrium_SI}A for 
convenience. 
Because the transitions between the states of the enzyme are not coupled to external energy consuming processes,
the choice of the rate constants is constrained by the cycle condition which states that the products of rate constants
in the clockwise and counterclockwise directions should be equal to each other in all the loops of the network diagram.
\cite{Hill2012}
Imposing the cycle condition results in the constraint equations for the different loops shown in 
Figure~\ref{fig:enzyme_equilibrium_SI}B.
In our analysis, we choose substrate unbinding to be the only process whose rate is different between right and wrong
substrates ($\koffW > \koffR$).
Therefore, the rate constants of all other identical processes are chosen to be the same between the substrates, i.e.
\begin{align}
\kAR=\kAW &\equiv \kAS, \\
\kIR=\kIW &\equiv \kIS, \\
\kARL=\kAWL &\equiv \kASL, \\
\kIRL=\kIWL &\equiv \kISL,
\end{align}
where the superscript ``S'' stands for both right (``R'') and wrong (``W'') substrates.
Using this general notation, we write the full set of constraint conditions on the rate constants obtained
from the different loops (mid, front, and back) as
\begin{align}
\label{eqn:constraint_mid}
\text{mid:} & \quad \frac{\kIL}{\kAL} = \frac{\loffA/\lonA}{\lonI/\loffI} \frac{\kI}{\kA}, \\
\text{front:} & \quad \frac{\kIS}{\kAS} = \frac{\konI}{\konA} \frac{\kI}{\kA}, \\
\label{eqn:constraint_back}
\text{back:} & \quad \frac{\kISL}{\kASL} = \frac{\konI}{\konA} \frac{\kIL}{\kAL}.
\end{align}
Note that the conditions imposed on the side loops follow directly from those of the other three loops via
\begin{align}
\text{side} = \frac{\text{mid} \times \text{back}}{\text{front}},
\end{align}
which is why the sides loop do not contribute a unique condition.

When writing the cycle conditions we did not include the product formation rate constant $r$, despite the
fact that production takes the enzyme into its substrate-free state, just like what unbinding through $\koff$ reactions does.
The reason for this is that $r$ is an effective rate constant for the process 
$\text{E} : \text{S} \xrightarrow{r} \text{E} + \text{P}$ representing the coarse-grained version of the full biochemical pathway of enzymatic production, namely,
$\text{E}:\text{S} \rightleftarrows \text{E}:\text{P} \rightarrow \text{E} + \text{P}$, which is distinct from the $\koff$ pathway of emptying the enzyme.
In our treatment we assume that product formation is practically irreversible which will be true if the product
concentration is kept low and, optionally, if the reverse reaction 
$\text{P} \rightarrow \text{S}$ is energetically highly unfavorable (e.g. 
requires a spontaneous formation of a covalent bond).

If the product formation rate is nonzero ($r>0$), the enzyme will be out of equilibrium despite the fact that its
individual transitions are not coupled to an energy source. This is due to the implicit assumption of having
the right and wrong substrate concentrations fixed, which makes the system open (i.e. new substrates enter
and products exit the system). 
We discuss the implications of this open system feature on the fidelity of the enzyme in the absence of driving
in the next section.

\newpage
\subsection{Enzyme Fidelity at a Fixed Ligand Concentration}
\label{section:SI_enzyme_equil_range}
As mentioned in the previous section, the presence of a nonzero production rate ($r>0$) makes the system open and
thereby takes the enzyme
out of equilibrium even at a fixed ligand concentration where the engine-enzyme coupling is absent.
For Hopfield's scheme, it can be shown
that in an analogous situation where driving is absent but the system is open,
the ``equilibrium'' (un-driven) fidelity is confined in a range defined by the ratio of the Michaelis and dissociation constants, which, 
for equal on-rates ($\kon^\text{R} = \kon^\text{W}$), becomes
\begin{align}
\label{eqn:eta_eq_range}
\eta_\text{eq} \in \left[ \frac{\koffW + r}{\koffR+r}, \frac{\koffW}{\koffR} \right].
\end{align}

\begin{figure*}[!ht]
	\centerline{
		\includegraphics[scale=1.00]{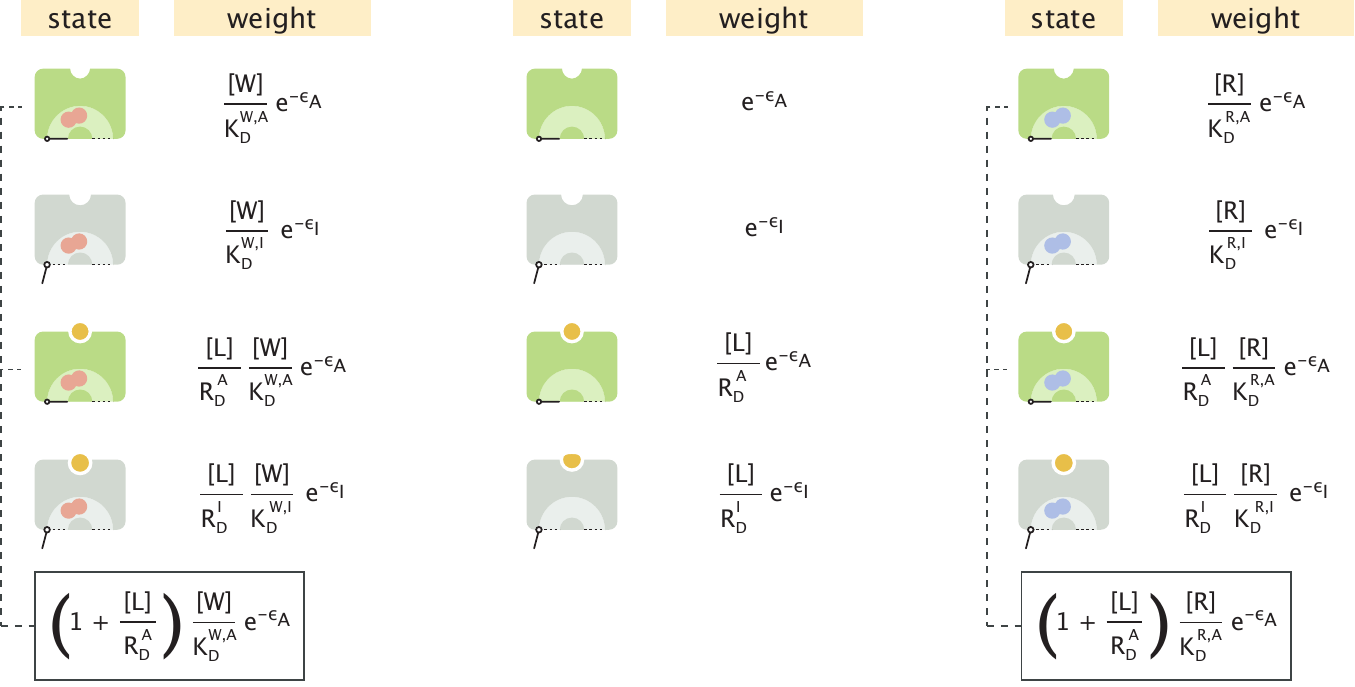}
	}
	\caption{
	Table of possible enzyme states and their statistical weights in the $r \rightarrow 0$ limit where the system
	is effectively at equilibrium. Here $\epsilon_\text{A}$ and $\epsilon_\text{I}$ stand for the energies of the 
	enzyme in its active and inactive states, respectively.
	The dissociation constants of the ligand and the substrates are denoted by 
	$R_\text{D}$ and $K_\text{D}$, respectively.
	}
	\label{fig:enzyme_states_weights}
\end{figure*}

We hypothesize that the same holds true for the allosteric enzyme as well despite the much wider
diversity of states available to it.
To demonstrate that, we first consider the limiting $r \rightarrow 0$ case where the product formation is so slow that 
the system effectively exists in a thermodynamic equilibrium. All possible enzyme states along with their statistical 
weights in this equilibrium setting are shown in Figure~\ref{fig:enzyme_states_weights}.
Fidelity can be found by adding the statistical weights of the right and wrong substrate-bound active states and dividing them, yielding
\begin{align}
\label{eqn_si:eta_kD}
\eta_\text{eq}(r \rightarrow 0) = 
\frac{[\text{R}]}{[\text{W}]} \frac{K_\text{D}^\text{W,A}}{K_\text{D}^\text{R,A}} = \frac{\koffW}{\koffR},
\end{align}
where we used $[\text{R}] = [\text{W}]$ and the equal on-rate assumption to go from dissociation constants to unbinding rates.
This corresponds to the upper limit in eq~\ref{eqn:eta_eq_range}.

\begin{figure*}[!ht]
	\centerline{
		\includegraphics[scale=1.00]{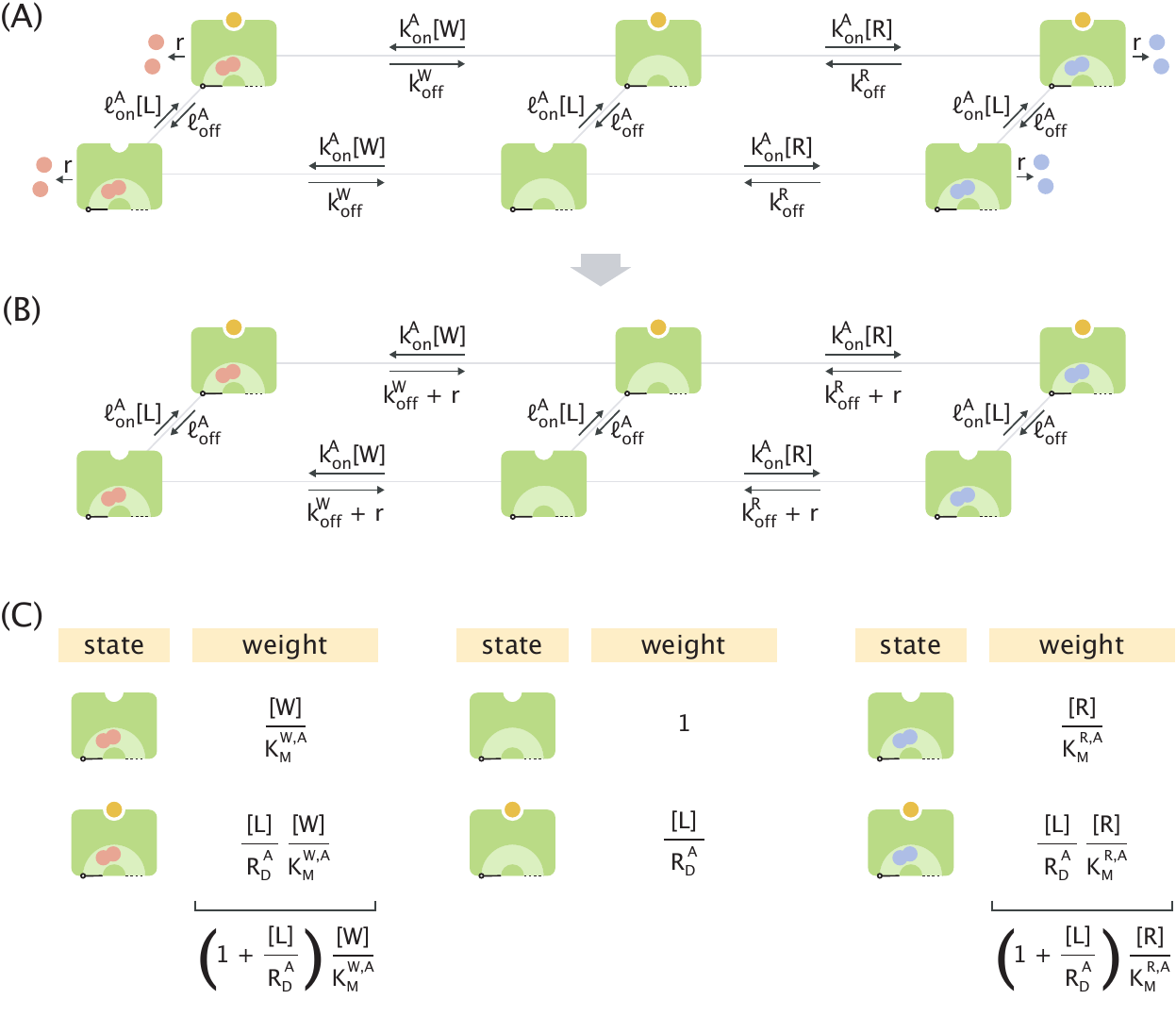}
	}
	\caption{
	The enzyme in the limit of constant activity and in the absence of engine coupling.
	(A) The active ``slice'' of the enzyme's full network diagram depicted in Figure~\ref{fig:enzyme_equilibrium_SI}A.
	(B) The reduced diagram corresponding to the network in panel A 
	with the production and off-rates combined under the same reaction arrow.
	(C) Table of the different enzyme states and their effective statistical weights.
	$K_\text{M}$ stands for the Michaelis constant.
	Total weights of the wrong and right substrate-bound states are shown below the left and right columns,
	respectively.
	}
	\label{fig:SI_open_limiting_case}
\end{figure*}

Intuitively, the presence of a nonzero production rate ($r>0$) should reduce the fidelity since the enzyme would have less time
to perform substrate filtering in its active state before product formation takes place.
To study how large this reduction can be, let us first consider a limiting case where the enzyme is exclusively in its active state
-- a setting where we expect the reduction effect to be manifested the most.
The active ``slice'' of the full network diagram corresponding to this limiting case is depicted in Figure~\ref{fig:SI_open_limiting_case}A.
Since product formation is just another path to substrate unbinding,
we can derive a corresponding reduced network diagram by adding
the production rate to the off-rates, as shown in Figure~\ref{fig:SI_open_limiting_case}B.

A peculiar feature of this network is that the cycle condition holds in its two loops, despite the fact
that the system is open ($r>0$).
This means that at steady state the detailed balance condition will hold
on all edges of the network (cf. Schnakenberg\cite{Schnakenberg1976}, section X),
allowing us to assign effective statistical mechanical weights to the different states
(Figure~\ref{fig:SI_open_limiting_case}C).

Dividing the total weights of the right and wrong substrate-bound states, we obtain the fidelity in this special
limit where the enzyme is exclusively in its active state, namely,
\begin{align}
\label{eqn_si:eta_kM}
\eta^\text{active}_\text{eq} (r > 0) = \frac{[\text{R}]}{[\text{W}]} \frac{K_\text{M}^\text{W,A}}{K_\text{M}^\text{R,A}} = \frac{\koffW+r}{\koffR+r}.
\end{align}
Here we again used $[\text{R}] = [\text{W}]$ and the equal on-rate assumption.
Note that this corresponds to the lower fidelity limit in the un-driven Hopfield model (eq.~\ref{eqn:eta_eq_range}).

We now hypothesize that the enzyme's fidelity falls between these two limits in the general case where the system is open ($r>0$)
and when the states are not constrained to be in the active ``slice'' of the full network diagram. 
Since obtaining the exact expression of fidelity in the general case is highly complicated due to the presence of a large number of states and loops in the network diagram,
and since a paper-and-pencil approach where the symmetries existing between the left and right ``wings'' of the network could potentially be taken advantage of to provide an analytical proof is also
not straightforward, we use a numerical method instead to justify our hypothesis.

\begin{figure*}[!ht]
	\centerline{
		\includegraphics[scale=1.00]{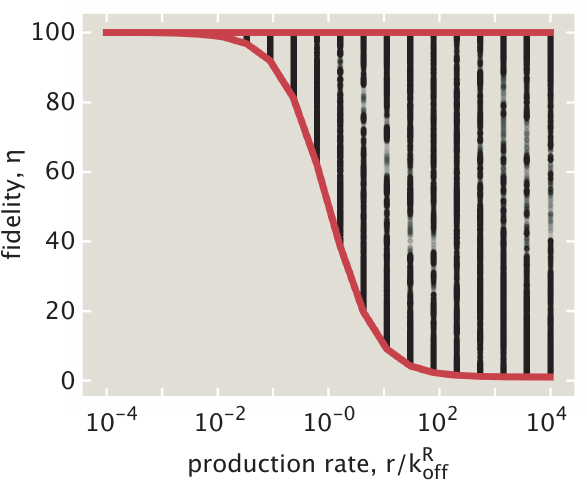}
	}
	\caption{
	Allosteric enzyme's fidelity in the absence of engine coupling (fixed $[\text{L}]$) for different choices of transition rates.
	The upper and lower red curves correspond to the ratios of the dissociation (eq~\ref{eqn_si:eta_kD})
	and Michaelis constants (eq~\ref{eqn_si:eta_kM}),
	respectively. Only the data points with fidelity values different up to the third significant digit were used 
	in the plot.	}
	\label{fig:SI_equil_enzyme_limits}
\end{figure*}

To that end, we fixed the ratio of the wrong and right substrate off-rates to be ${\koffW/\koffR=100}$, sampled values for 
enzyme's remaining
transition rate constants from the $[10^{-4}  \koffR, 10^4  \koffR]$ range (generating 20,971,520 independent sets in total),
and evaluated the fidelity for each parameter set.
The results of the numerical study are summarized in Figure~\ref{fig:SI_equil_enzyme_limits}.
As can be seen, despite the wider diversity of allosteric enzyme's states, its fidelity in the absence of engine coupling
still falls between the ``equilibrium'' limits of Hopfield's model (eq~\ref{eqn:eta_eq_range}).

\newpage
\section{Full Description of the Piston Model with Engine-Enzyme Coupling}

In this section, we provide details on the analytical and numerical explorations of the full model.
In \ref{section:SI_coupled_equil} we discuss the thermodynamics of coupling the engine to the allosteric enzyme.
Then, in \ref{section:SI_p_ss} we present the methodology for obtaining the steady state probabilities of system states 
under external drive.
In \ref{section:param_values} and \ref{section:global_optimization_details} 
we provide the parameters used in the numerical study of section 3.3 and
describe the fidelity optimization strategy used in study of the section 3.4, respectively.
Lastly, in \ref{section:alpha3_investigate} we
investigate in detail the $\alpha_\text{max} \approx 3$ result for the highest fidelity of the model.

\subsection{Equilibrium Fidelity of the Piston Model in the Absence of External Driving}
\label{section:SI_coupled_equil}

In Supporting Information section \ref{section:SI_enzyme_equil_range} 
we showed that at a fixed ligand concentration the fidelity
of the allosteric enzyme was constrainted within the range given in eq~\ref{eqn:eta_eq_range}.
Here we demonstrate that the same result holds also for the full model in the absence of external driving
when a thermodynamically consistent coupling is made between the engine and the enzyme.

\begin{figure*}[!ht]
	\centerline{
		\includegraphics[scale=1.00]{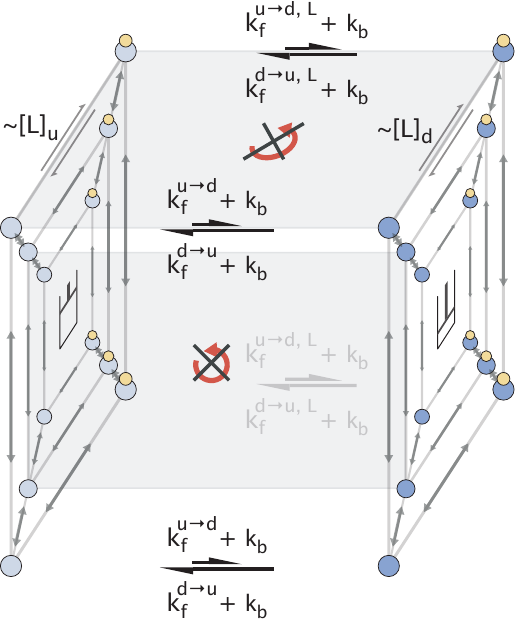}
	}
	\caption{
	The effective network diagram of the piston model in the absence of driving.
	The forward and backward pathways connecting the two layers of the diagram are combined to yield effective rates.	
	The two kinds of cycles where ligand binding events are present or absent are shown as horizontal and vertical shaded
	rectangles, respectively.
	The crossed cycling arrows indicate the absence of driving forces in the shaded loops.
	}
	\label{fig:SI_coupling_equil}
\end{figure*}

In the absence of driving, the finite-state equivalent of the full network (Figure 5D) can be
reduced into the one shown in Figure~\ref{fig:SI_coupling_equil}
where we have combined the ratchet transitions through forward and backward
pathways under a single arrow -- a procedure allowed when the transitions are not driven.\cite{HorowitzPRE2017}
Because of the equilibrium constraints imposed on the enzyme's transition rates discussed in Supporting Information
section~\ref{section:enzyme_cycle},
the cycle condition will hold for the loops in the left and right ``layers'' of the diagram in Figure~\ref{fig:SI_coupling_equil}.
The loops where the cycle condition could possibly be violated are the ones that involve transitions between the two layers,
i.e. piston compressions and expansions.

The first class of such loops does not involve ligand binding and unbinding events (for example,  the shaded vertical rectangle in
Figure~\ref{fig:SI_coupling_equil}) and therefore, 
the cycle condition is automatically satisfied in such loops since for each clockwise transition there is a corresponding counterclockwise transition with an identical rate.
The second class of loops that connect the two layers involves ligand binding and unbinding events which affect the 
rate of switching between the layers (e.g. the shaded horizontal rectangle in Figure~\ref{fig:SI_coupling_equil}).
The driving force in these loops is given by
\begin{align}
\label{eqn:dmu_coupled_eq}
\Delta \mu = \beta^{-1} \ln \frac{( \kfud + \kb ) ( \kfduL + \kb ) [\text{L}]_\text{d}}
{( \kfdu + \kb ) ( \kfudL + \kb ) [\text{L}]_\text{u}},
\end{align}
where we used the fact that the ligand binding rates are proportional to ligand concentrations. 
Now, in the general case where there are $N$ ligands under the piston (one of which can be bound to the enzyme),
the different forward stepping rates become
\begin{align}
\kfud &= \kb \text{e}^{-\beta (\Delta W_\text{eq} +  \beta^{-1} N \ln f )} = \kb f^{-N}, \\
\kfudL &= \kb \text{e}^{-\beta (\Delta W_\text{eq} +  \beta^{-1} (N-1) \ln f)} = \kb f^{-(N-1)}, \\
\kfdu &= \kb \text{e}^{-\beta (\Delta W_\text{eq} -  \beta^{-1} N \ln f  )} = \kb f^N, \\
\kfduL &= \kb \text{e}^{-\beta (\Delta W_\text{eq} -  \beta^{-1} (N-1) \ln f)} = \kb f^{N-1}.
\end{align}
Here we set $\Delta W_\text{eq} = 0$ to account for the absence of driving and used the fact that the free energy change of
$N$ ligands upon isothermal compression is $\beta^{-1} N \ln f$ (with a negative sign upon expansion) and that $N$ should be
replaces with $N-1$ when one of the ligands is bound to the enzyme.

Substituting these expressions into eq~\ref{eqn:dmu_coupled_eq} and using the identity $[L]_\text{d} = f[L]_\text{u}$, we find
\begin{align}
\Delta \mu &= \beta^{-1} \ln \frac{(f^{-N} + 1) (f^{N-1} + 1) f}{(f^N + 1)(f^{1-N} + 1)} \nonumber\\
&= \beta^{-1} \ln \left( \frac{f^{-N} (1 + f^N)}{f^N + 1} \times \frac{f^{N-1} (1 + f^{1-N})}{f^{1-N} + 1} \times f \right) \nonumber\\
&= \beta^{-1} \ln \left( f^{-N} \times f^{N-1} \times f \right) \nonumber\\
&= \beta^{-1} \ln 1 = 0.
\end{align}
This shows that in the absence of external driving ($\Delta W = 0$) the cycle condition holds for all loops of the network,
demonstrating the thermodynamic consistency of the coupling between the engine and the enzyme.

As in our separate treatment of the allosteric enzyme in Supporting Information section~\ref{section:SI_enzyme_equil_range},
here too in the $r \rightarrow 0$ limit the system will approach a thermodynamic equilibrium.
Since we already know that in the equilibrium limit the fidelity of the enzyme at a fixed ligand concentration is given by the
ratio of the wrong and right off-rates, we can apply this result to the left and right layers of the diagram in 
Figure~\ref{fig:SI_coupling_equil} and obtain a relation between the net statistical weights of the
right and wrong substrate-bound active states, namely,
\begin{align}
\frac{w^\text{R}_\text{u}}{w^\text{W}_\text{u}} = \frac{w^\text{R}_\text{d}}{w^\text{W}_\text{d}} = \frac{\koffW}{\koffR}.
\end{align}
Here ``u'' and ``d'' stand for the expanded (left layer) and compressed (right layer) states of the piston. 
We can then write the fidelity of the full network in terms of these weights as
\begin{align}
\eta_\text{eq} (r \rightarrow 0) &= \frac{w^\text{R}_\text{u} + w^\text{R}_\text{d}}{w^\text{W}_\text{u} + w^\text{W}_\text{d}} 
= \frac{w^\text{W}_\text{u} \left(\frac{\koffW}{\koffR}\right) + w^\text{W}_\text{d} \left(\frac{\koffW}{\koffR}\right)}{w^\text{W}_\text{u} + w^\text{W}_\text{d}} = \frac{\koffW}{\koffR},
\end{align}
which corresponds to the upper limit of the equilibrium fidelity range in eq~\ref{eqn:eta_eq_range}.

\begin{figure*}[!ht]
	\centerline{
		\includegraphics[scale=1.00]{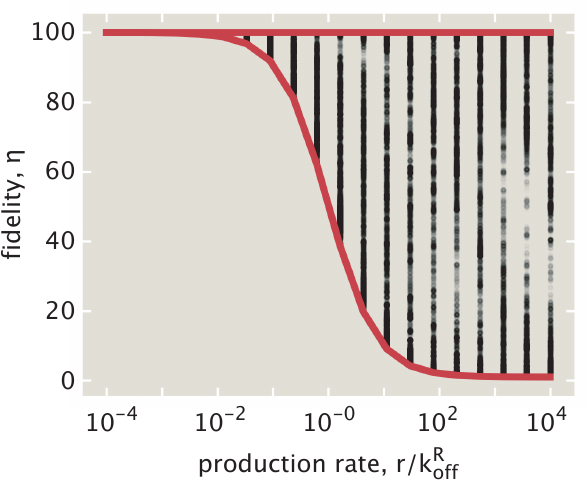}
	}
	\caption{
	Fidelity of the full piston model in the absence of driving ($\Delta W = 0$) for different choices of model parameters.
	The upper and lower red curves correspond to the ratios of the dissociation (eq~\ref{eqn_si:eta_kD})
	and Michaelis constants (eq~\ref{eqn_si:eta_kM}),
	respectively.
	Only the data points with fidelity values different up to the third significant digit were used 
	in the plot.
	}
	\label{fig:SI_system_equil}
\end{figure*}

To demonstrate that in the absence of driving the coupled system meets also the lower fidelity limit given by 
$(\koffW+r)/(\koffR+r)$, we again use a numerical approach
and sample the parameter space, evaluating the fidelity at each of the 10,628,820 sampled sets of parameters.
As in the study of Figure~\ref{fig:SI_coupling_equil}, 
here too we set the ratio of off-rates to be $\koffW/\koffR = 100$.
The results are summarized in Figure~\ref{fig:SI_system_equil}, where it can be seen that all
points lie between the limits of eq~\ref{eqn:eta_eq_range}. 
Overall, this study shows that in the absence of driving, the coupling of the engine to the enzyme alone cannot
lead to fidelity enhancement.

\newpage
\subsection{Obtaining the Steady-State Occupancy Probabilities}
\label{section:SI_p_ss}

\begin{figure*}[!ht]
	\centerline{
		\includegraphics[scale=1.00]{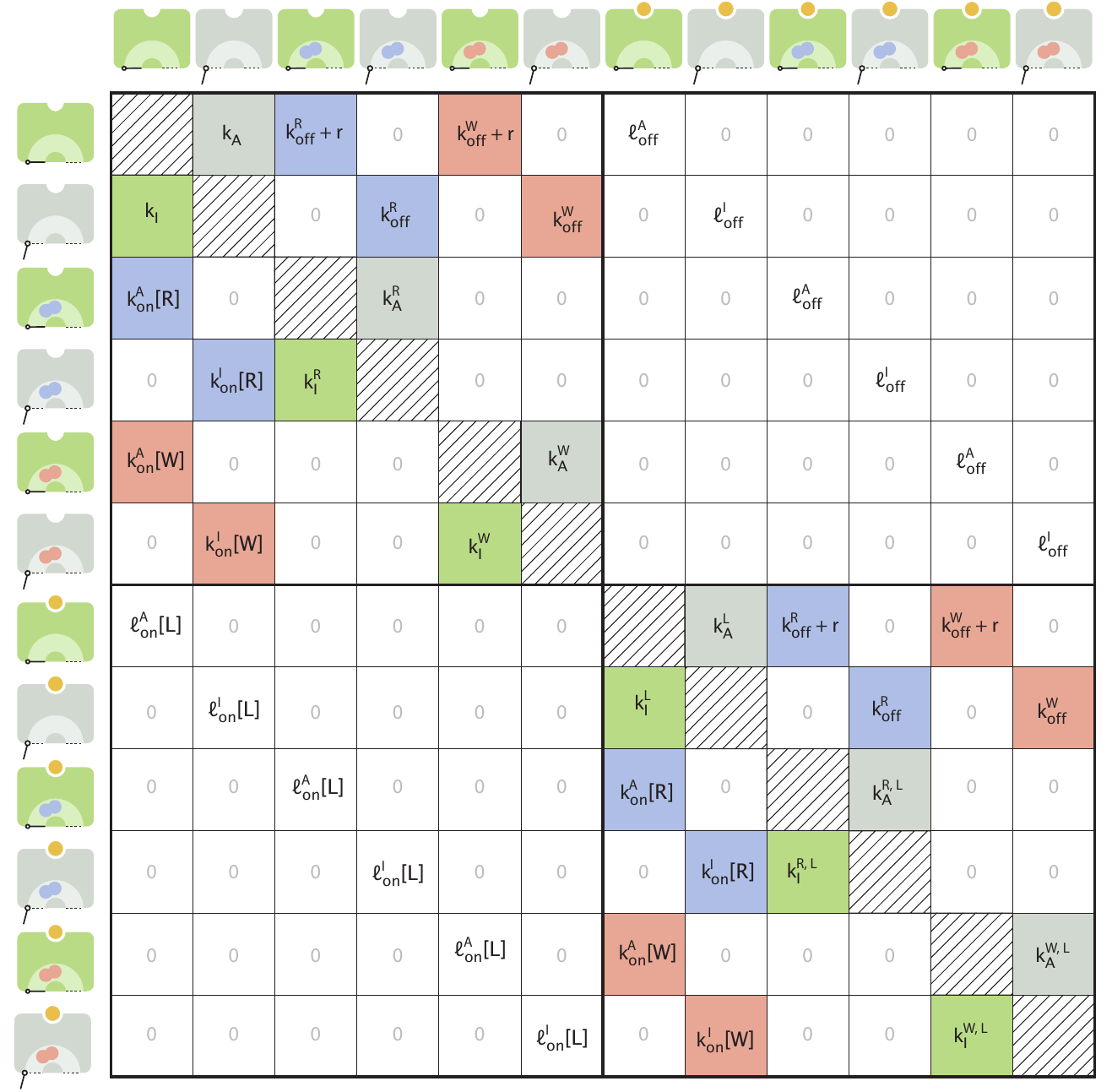}
	}
	\caption{Transition rates between the enzyme states. The element $Q^\text{enzyme}_{i,j}$
	stands for the rate of transitioning from the $j^\text{th}$ into the $i^\text{th}$ state of the enzyme ($i \ne j$)
	at a fixed ligand concentration $[L]$. 
	Red- and blue-colored cells show transitions involving the binding or release of incorrect and correct
	substrates, respectively. Green- and gray-colored cells show inactivating and activating enzyme transitions, respectively.
	The diagonals are shaded to indicate that they are not used when constructing $\Qb$.
	}
	\label{fig:transitions_12_states}
\end{figure*}

The kinetics of the full piston model is characterized by a $24\times24$ transition rate matrix $\mathbf{Q}$,
which has the block form
\begin{align}
\Qb = 
\begin{pmatrix}
\vspace{1mm}
\Qb^\text{enzyme}_\text{u} & \Qb_{\text{d} \rightarrow \text{u}} \\
\vspace{1mm}
\Qb_{\text{u} \rightarrow \text{d}} & \Qb^\text{enzyme}_\text{d}
\end{pmatrix}.
\end{align}
Here the non-diagonal elements of the $12\times12$ matrices 
$\Qb^\text{enzyme}_\text{u}$ and $\Qb^\text{enzyme}_\text{d}$ represent the
transition rates between the different enzyme states when the ligand concentration is $[L]=[L]_\text{u}$
and $[L]=[L]_\text{d}$, respectively. These non-diagonal terms at a given ligand concentration $[L]$
are depicted in Figure~\ref{fig:transitions_12_states}.

The other two matrices, namely, $\Qb_{\text{d} \rightarrow \text{u}}$ and $\Qb_{\text{u} \rightarrow \text{d}}$,
are diagonal whose elements stand for the net piston compression ($\text{d} \rightarrow \text{u}$) and 
expansion ($\text{u} \rightarrow \text{d}$) rates that alter the state of the piston but leave the state of the enzyme
unchanged. They are given by
\begin{align}
\Qb_{\text{d} \rightarrow \text{u}} &=
\text{diag} \bigg(
\underbrace{ \big( \kb + \kfdu \big), ..., \big(\kb + \kfdu \big)}_{\text{6 terms}}, 
\underbrace{ \big( \kb + \kfduL \big), ..., \big(\kb + \kfduL \big)}_{\text{6 terms}}
\bigg), \\
\Qb_{\text{u} \rightarrow \text{d}} &=
\text{diag} \bigg(
\underbrace{ \big( \kb + \kfud \big), ..., \big(\kb + \kfud \big)}_{\text{6 terms}}, 
\underbrace{ \big( \kb + \kfudL \big), ..., \big(\kb + \kfudL \big)}_{\text{6 terms}}
\bigg).
\end{align}
Note that since the forward stepping rates depend on whether the ligand is bound or not,
they appear without a superscript ``L'' in the first 6 terms (where the ligand unbound),
and with a superscript ``L'' in the last 6 terms (where the ligand is bound). 
Lastly, the diagonal elements of $\Qb$ are assigned such that $Q_{jj} = -\sum_{i\ne j} Q_{ij}$,
ensuring that the columns sum to zero.

The dynamics of the coupled engine-enzyme system is described via
\begin{align}
\frac{\mathrm{d} \vec{p}}{\mathrm{d} t} = \Qb \, \vec{p},
\end{align}
where $\vec{p}$ is a column vector whose 24 elements stand for the probabilities of the different system states
(12 enzyme states $\times$ 2 piston states).
We are interested in the steady state behavior of the piston model, where $\mathrm{d} \vec{p} / \mathrm{d}t = \vec{0}$.
Since the exact analytical expressions for the steady state probabilities $\vec{p}_\text{ss} \equiv \vec{\pi}$ are highly convoluted,
in our parametric studies 
we use numerical methods to find $\vec{\pi}$ from $\Qb \vec{\pi} = \vec{0}$ and $\sum_i \pi_i = 1$,
where the latter equation guarantees that the probabilities sum to 1.

\newpage
\subsection{Enzyme's Kinetic Parameters Used for the Numerical Study in Section 3.4}
\label{section:param_values}

Here we provide the list of enzyme's transition rates used for numerically studying the effects of tuning the engine
``knobs'' in section 3.4.
Since none of the performance metrics used in the study depend on the absolute timescale of the model's dynamics,
we set the unbinding rate of the right substrate to be unity ($\koffR = 1$),
and defined all other rates relative to it. Specifically, we chose $\koffW = 100$ so that the fidelity after a single proofreading realization roughly matched that
of the ribosome ($\eta_\text{translation} \sim 10^4$).\cite{Hopfield1974} 
Also, we chose the catalysis rate to be much slower compared
with the off-rates ($r = 0.2$) - a condition for high fidelity suggested in Hopfield's original paper.

The remaining rate constants were assigned values that meet the intuitive expectations from the conceptual
introduction of the model in section 2. 
Specifically, the rate of substrate binding to the active enzyme was chosen to be much less than the rate
of binding to the inactive enzyme in order to yield low leakiness ($\konA/\konI = 10^{-5} \ll 1$).
Next, the enzyme was chosen to be predominantly inactive in its native state to allow for new substrate binding events
($\kI / \kA = 50 \gg 1$).
Lastly, the rates of ligand binding and unbinding were assigned values that ensure that the ligand acts as a strong
activator ($ \frac{\loffI / \lonI}{\loffA / \lonA} = 10^6 \gg 1$).

The values of the independent parameters $\kAL$ and $\kAS$ were assigned after manually inspecting the effect of
different numerical choices on the model performance.
Finally, the values of the remaining four parameters (lower section in Table~\ref{table:parameter_values}) 
were calculated from the cycle conditions in eqs~\ref{eqn:constraint_mid}-\ref{eqn:constraint_back}
under the assumption that ligand binding does not alter the ratio of inactivation or activation rates
in the substrate-bound and substrate-unbound states (i.e. $\kISL/\kIL = \kIS/\kI$ and $\kASL/\kAL = \kAS/\kA$).

\begin{table}[h!]
\centering
\begin{tabular}{| p{5 cm} | p{3cm}|}
\hline
Transition rate & Value \\
\hline
\hline
$\koffR$, $\koffW$ & $1$, $100$ \\
$r$ & $0.2$ \\
$\konA [\text{S}]$ & $10^{-5}$\\
$\konI [\text{S}]$ & $1$\\
$\kA$ & $20$\\
$\kI$ & $1000$\\
$\lonA$ & $0.1$\\
$\loffA$ & $5$\\
$\lonI$ & $0.01$\\
$\loffI$ & $500000$\\
$\kAL$ & $2000$\\
$\kAS$ & $0.01$\\
\hline
\hline
$\kIL$ & $0.1$ \\
$\kIS$ & $50000$ \\
$\kASL$ & $1$ \\
$\kISL$ & $5$ \\
\hline
\end{tabular}
\caption{Values of enzyme's different transition rates used in the studies of Figure 6.}
\label{table:parameter_values}
\end{table}

\newpage
\subsection{Details of the Numerical Optimization Procedure for Finding the Highest Fidelity}
\label{section:global_optimization_details}

In our optimization scheme, we first chose the values of rates which were kept fixed for the rest of the study.
These include the unbinding rate of right substrates ($\koffR = 1$), the catalysis rate ($r=0.2$), and the effective
first-order rate of substrate binding to the inactive enzyme state ($\konI [\text{S}] = 1$).
Also, since no limits were imposed on the amount of energy expenditure, 
we chose large values for the compression factor ($f=10^{100}$) and 
the work per step ($\Delta W = 1000 \, k_\text{B}T$) to maximize the quality of proofreading.

Then, we considered a set of 144 different
initialization options for the remaining parameters
to be used in our numerical optimization procedure.
To avoid the completely independent tuning of related enzyme activation and inactivation rates, we
considered three possible options that met the cycle condition. Namely, 1) $\kAS = \kA$ and $\kIS = \kI /  \gamma$,
2) $\kAS =  \sqrt{\gamma} \kA$ and $\kIS = \kI / \sqrt{\gamma} $, 3) $\kAS = \gamma \kA$ and $\kIS = \kI$, where
$\gamma = \konA/\konI$. All of these three options satisfy the cycle constraint $\konA \kIS \kA = \konI \kAS \kI$ (Figure~\ref{fig:enzyme_equilibrium_SI}B). 
Options for the transition rates between ligand-bound enzyme states (i.e. $\kASL$ and $\kISL$)
were chosen analogously.

\begin{figure*}[!ht]
	\centerline{
		\includegraphics[scale=1.00]{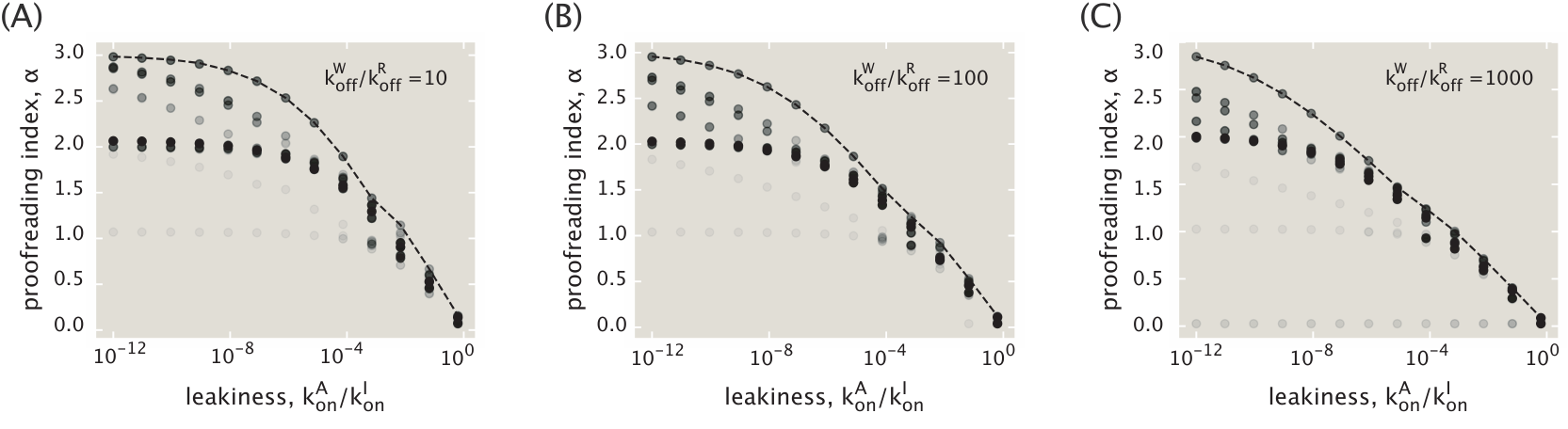}
	}
	\caption{
	Fidelity optimization results for each of the 144 parameter initialization options.  
	(A) $\koffW/\koffR = 10$. (B) $\koffW/\koffR = 100$. (C) $\koffW/\koffR = 1000$.
	The dotted lines in each panel represent the trends for the globally optimal fidelities.
	}
	\label{fig:si_global_optimization}
\end{figure*}

In our custom-written maximization algorithm
we iteratively perturbed all the parameters for multiple rounds with decreasing amplitudes
until the convergence criterion was met or until the number of iterations exceeded the specified threshold
(at most 20 iterations for each of the 6 decreasing amplitudes).
The results from each of these local maximization procedures are summarized in Figure~\ref{fig:si_global_optimization}.
We chose the largest among the different local maxima to represent the highest fidelity available for the
given $\left( \konA, \koffW \right)$ pair.

\newpage
\subsection{Investigation of the $\alpha_{\max} \approx 3$ Result for the Highest Available Proofreading Index}
\label{section:alpha3_investigate}

\begin{figure*}[!ht]
	\centerline{
		\includegraphics[scale=1.00]{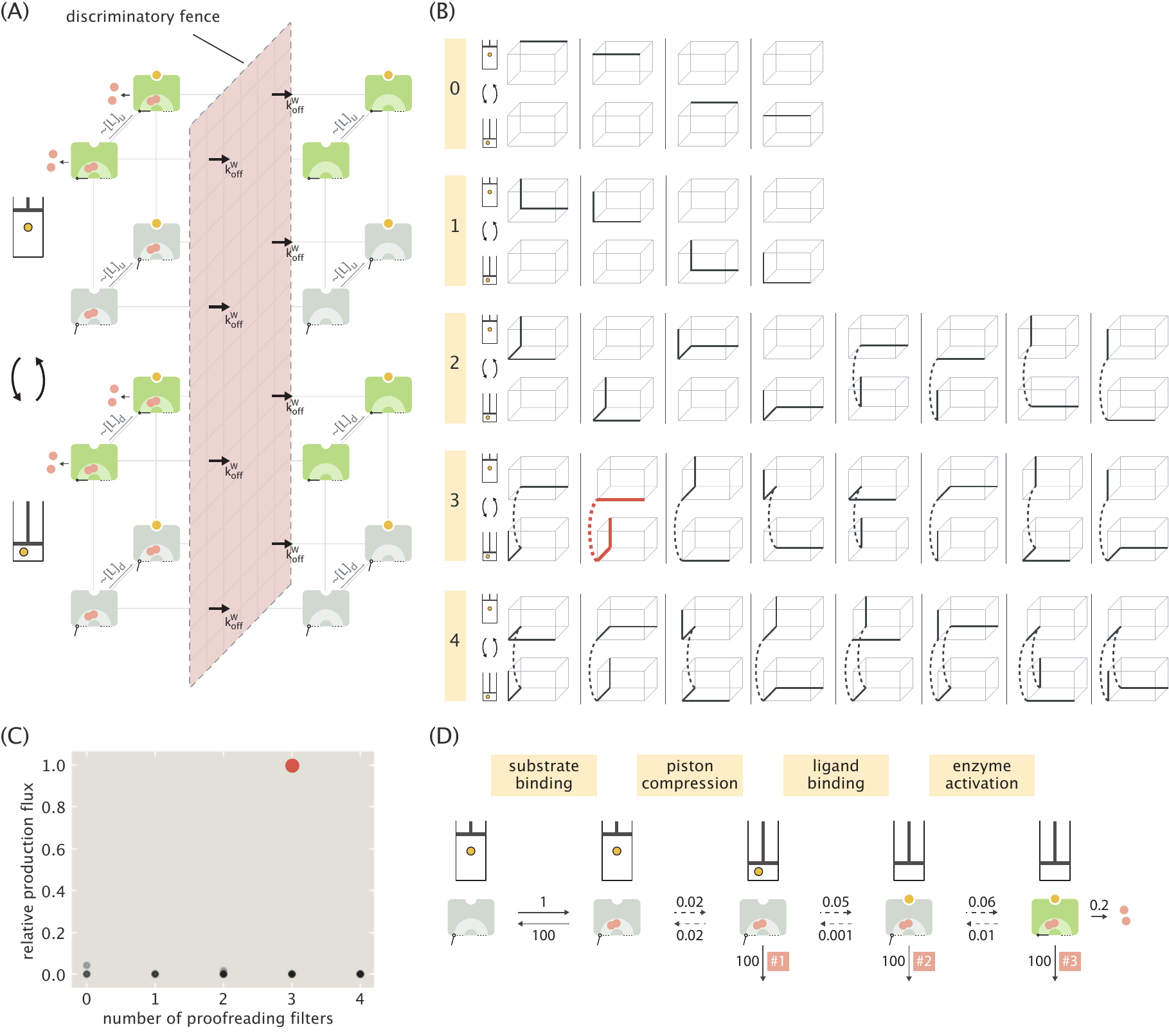}
	}
	\caption{
	(A) The wrong ``wing'' of the full reaction network along with the discriminatory fence. Ligand concentrations
	that enter the ligand binding rates are shown to indicate the difference between the upper and lower halves
	of the diagram. (B) The full set of unique trajectories that start on the right side of the network and end up 
	at an active enzyme state on the left side.
	Numbers of proofreading filters available to trajectories are shown
	on the side. 
	Piston state transitions are marked with dotted lines for clarity.
	The dominant trajectory in panel (C) is highlighted in red.
	(C) The relative product formation fluxes of all possible trajectories calculated for the 
	case where $\koffW/\koffR = 100$ and $\konA/\konI = 10^{-12}$, and grouped by the number of filters. The red
	dot indicates the dominant one. (D) Schematic illustration of the dominant trajectory in panel (C) along with 
	the numerical values of the rates. The dotted arrows suggest that the intermediate transitions
	are much slower than the off-rate.
	}
	\label{fig:si_wall}
\end{figure*}

Our numerical scheme for optimizing the fidelity (Supporting Information section~\ref{section:global_optimization_details})
revealed that the piston model could perform proofreading up to three times ($\alpha_\text{max}\approx3$).
To gain intuition on how this is possible, let us consider the wrong ``wing'' of the full reaction network
(Figure~\ref{fig:si_wall}A). Each system state is characterized by the piston position (up or down) and the state of the
enzyme (one of 8 possibilities).
To turn a wrong substrate into a product, the system needs to traverse a trajectory that starts 
at a substrate-unbound state 
on the right side of the
diagram and reaches one of the substrate-bound active states on the left side,
at which point catalysis can take place.
Using the terminology introduced in Murugan, \textit{et al.}\cite{MuruganPRX2014}, we can say that a proofreading
filter can be realized every time the system makes a transition parallel to the ``discriminatory fence'' of the network (Figure~\ref{fig:si_wall}A). Rates which are on either side of the fence do not discriminate between the
two kinds of substrates; only those that cross the fence do, which in our case are the off-rates ($\koffW > \koffR$).
Thus, the number of such parallel transitions that the system makes before reaching the catalytically active state
represents the largest number of proofreading filters available to the given trajectory.

Figure~\ref{fig:si_wall}B shows the full list of unique trajectories that start on the right side of the network, cross the
discriminatory fence and eventually reach an active enzyme state after traversing through a series of
inactive states. The trajectories are grouped by the number of these inactive states visited on the left side of the wing
prior to reaching the active state.
For example, entries of the first group represent trajectories where the substrate binds directly
to the active enzyme and hence, undergoes zero proofreading filtrations. The discriminatory capacity of the 
piston model
will therefore depend on which of the trajectories dominates in product formation.

To compare the contributions from different trajectories, we assign each of them a probability flux, which approximates
the average rate of product formation through the trajectory. We define this flux via
\begin{align}
\label{eqn:flux}
J_{\vec{s}} = \pi_{s_0} k_{s_0 \rightarrow s_1} \left( \prod_{i=1}^{N-2} p_{s_i \rightarrow s_{i+1}} \right)  p^\text{cat}_{s_{N-1}},
\end{align}
where $\vec{s}$ is the set of $N$ states in the trajectory, 
$\pi_{s_0} k_{s_0 \rightarrow s_1}$ is the substrate binding flux that crosses the fence at the start of the trajectory,
$p_{s_i, s_{i+1}}$ are the probabilities of staying on the trajectory during traversal, 
and, lastly, $p^\text{cat}_{s_{N-1}}$ is the probability of catalysis once the system has reached the active enzyme state $s_{N-1}$.
Note that the flux expression does not account for backtracking events whose contribution we expect to be
insignificant for efficient proofreading trajectories, since for them $p_{s_i \rightarrow s_{i+1}} \ll 1$.

Having defined a flux metric for each trajectory, we then calculated its value for all trajectories listed in 
Figure~\ref{fig:si_wall}B in the case where $\koffW/\koffR = 100$ and $\konA/\konI = 10^{-12}$ (low leakiness).
Figure~\ref{fig:si_wall}C shows the fluxes normalized by the highest one and 
grouped by the number of proofreading filters.
As we can see, the dominant trajectory indeed contains three filters.
This dominant trajectory is highlighted in red in Figure~\ref{fig:si_wall}B and also corresponds to the one
shown in Figure 7B of the main text.

We would like to note here that the model parameters inferred from the unconstrained fidelity optimization 
were degenerate, and there was an alternative set with $\alpha \approx 3$ proofreading index whose
corresponding dominant trajectory was different from the one highlighted in Figure~\ref{fig:si_wall}B.
Some of the parameters of this set, however, contradicted our model criteria (e.g. the binding rate
in the expanded piston state was very high), which is why we did not use this alternative set for our main discussion.
Parameters that did satisfy our model criteria are shown for the $\koffW/\koffR = 100$ case in
Figure~\ref{fig:si_wall}D. The transition rates between intermediates are much slower compared
with the off-rate, as expected for an efficient proofreading performance.

Lastly, as can be seen in Figure~\ref{fig:si_wall}B, the highest number of filters that a unique
trajectory could, in principle, realize is 4 and not 3. This raises the question of why a trajectory with 4 potential
filters cannot be a dominant one, as our numerical results in Figure~\ref{fig:si_wall}C have suggested.
We answer this question for three representative cases and invite the reader to work through the remaining examples.
Our approach will be to show that the flux through a given 4-filter trajectory is necessarily less than that of
some other trajectory with fewer filters, which would suggest that it cannot be a dominant one.

\begin{figure*}[!ht]
	\centerline{
		\includegraphics[scale=1.00]{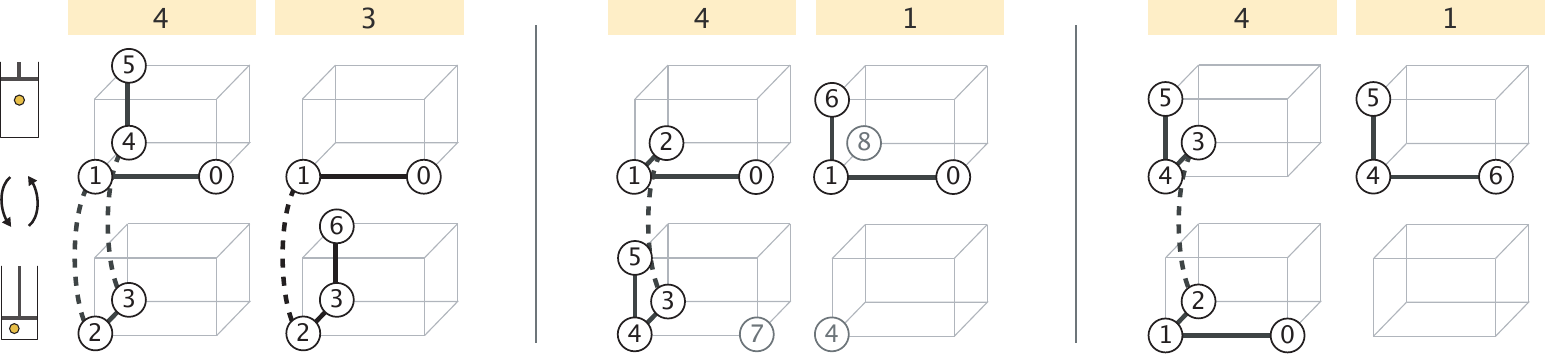}
	}
	\caption{Three representative 4-filter trajectories paired with ones which have a lower filter number and, 
	necessarily, a higher product formation flux. 
	The state indices are added to facilitate the comparison between the corresponding trajectories.
	}
	\label{fig:si_no_four_filter}
\end{figure*}

Figure~\ref{fig:si_no_four_filter} shows three different 4-filter trajectories 
next to corresponding trajectories with fewer filters,
flux through which, as we will show, will necessarily be greater.
Throughout our analysis we will be making use of the fact that the rates of
piston expansion and compression are identical (and equal to $\kb$)
in the large driving limit considered here.
Let us start from the first example. 
Using Eq~\ref{eqn:flux}, we can write the fluxes through 4-filter ($J_{(4)}$) and 3-filter ($J_{(3)}$) trajectories 
respectively as
\begin{align}
J_{(4)} &= \pi_0 k_{0 \rightarrow 1} \times p_{1 \rightarrow 2} p_{2 \rightarrow 3} p_{3 \rightarrow 4}
p_{4 \rightarrow 5} \times p^\text{cat}_{5}, \\
J_{(3)} &= \pi_0 k_{0 \rightarrow 1} \times p_{1 \rightarrow 2} p_{2 \rightarrow 3} p_{3 \rightarrow 6} \times p^\text{cat}_{6}.
\end{align}
Since the states (3) and (4) correspond to the same enzyme state and have identical outgoing rates, we have
$p_{4 \rightarrow 5} = p_{3 \rightarrow 6}$. From the same argument for states (5) and (6) we find 
$p^\text{cat}_5 = p^\text{cat}_6$. With these identities at hand, we can write the ratio of the two fluxes as
\begin{align}
\frac{J_{(4)}}{J_{(3)}} = p_{3 \rightarrow 4} < 1.
\end{align}
Therefore, the 4-filter trajectory is necessarily slower than the 3-filter one and cannot 
dominate the dynamics of wrong product formation.

Now let us look at the slightly more complicated second example. There the fluxes of the 4-filter and 1-filter
trajectories are
\begin{align}
J_{(4)} &= \pi_0 k_{0 \rightarrow 1} \times p_{1 \rightarrow 2} p_{2 \rightarrow 3} p_{3 \rightarrow 4}
p_{4 \rightarrow 5} \times p^\text{cat}_{5}, \\
J_{(1)} &= \pi_0 k_{0 \rightarrow 1} \times p_{1 \rightarrow 6} \times p^\text{cat}_{6},
\end{align}
respectively. The full expression of the transition probability $p_{4 \rightarrow 5}$ is
\begin{align}
p_{4 \rightarrow 5} = \frac{k_{4 \rightarrow 5}}{k_{4 \rightarrow 5} + k_{4 \rightarrow 3} + k_{4 \rightarrow 1} + 
k_{4 \rightarrow 7}}.
\end{align}
Similarly, the expression for $p_{1 \rightarrow 6}$ is
\begin{align}
p_{1 \rightarrow 6} = \frac{k_{1 \rightarrow 6}}{k_{1 \rightarrow 6} + k_{1 \rightarrow 8} + k_{1 \rightarrow 4} + 
k_{1 \rightarrow 0}}.
\end{align}
All corresponding rates in the above probability expressions are equal to each other
(i.e. $k_{4 \rightarrow 5} = k_{1 \rightarrow 6} = \kAS$, $k_{4 \rightarrow 1} = k_{1 \rightarrow 4} = \kb$, $k_{4 \rightarrow 7} = k_{1 \rightarrow 0} = \koffW$),
with the exception of $k_{4 \rightarrow 3} = \lonI [\text{L}]_\text{d}$ and $k_{1 \rightarrow 8} = \lonI [\text{L}]_\text{u}$.
Now, since $[\text{L}]_\text{d} > [\text{L}]_\text{u}$, we obtain $p_{4 \rightarrow 5} < p_{1 \rightarrow 6}$.
With an identical reasoning we can also find that $p^\text{cat}_{5} < p^\text{cat}_{6}$.
Therefore, the ratio of the 4-filter and 1-filter trajectory fluxes becomes
\begin{align}
\frac{J_{(4)}}{J_{(1)}} = p_{1 \rightarrow 2} p_{2 \rightarrow 3} p_{3 \rightarrow 4}
\underbrace{\left( \frac{p_{4 \rightarrow 5}}{p_{1 \rightarrow 6}} \right)}_{<1}
\underbrace{\left( \frac{p^\text{cat}_{5}}{p^\text{cat}_{6}} \right)}_{<1} < 1,
\end{align}
proving our claim.

Lastly, we consider the third example in Figure~\ref{fig:si_no_four_filter}.
We again start off by writing the trajectory fluxes, namely,
\begin{align}
J_{(4)} &= \pi_0 k_{0 \rightarrow 1} \times p_{1 \rightarrow 2} p_{2 \rightarrow 3} p_{3 \rightarrow 4}
p_{4 \rightarrow 5} \times p^\text{cat}_{5}, \\
J_{(1)} &= \pi_6 k_{6 \rightarrow 4} \times p_{4 \rightarrow 5} \times p^\text{cat}_{5}.
\end{align}
The two rates appearing in the flux expression represent the substrate binding rate and are equal to each other, that is,
$ k_{0 \rightarrow 1} =  k_{6 \rightarrow 4} = \konI [\text{S}]$. Now, note that $\pi_6$ is the steady state probability of the 
inactive ligand-unbound enzyme state at a low ligand concentration, whereas $\pi_0$ is the probability of the same
enzyme state at a high ligand concentration. Since these are ligand-unbound states, the one in the presence
of a lower ligand concentration will have a higher probability, i.e. $\pi_6 > \pi_0$. Thus, taking the ratio of the two
fluxes, we obtain
\begin{align}
\frac{J_{(4)}}{J_{(1)}} = \underbrace{\left( \frac{\pi_0}{\pi_6} \right)}_{<1} p_{1 \rightarrow 2} p_{2 \rightarrow 3} p_{3 \rightarrow 4} < 1,
\end{align}
suggesting that the 4-filter trajectory in the third example too cannot be the dominant one.

\newpage
\bibliography{PaperLibrary}